\RequirePackage{fix-cm}

\documentclass{svjour3}                     

\usepackage{graphicx}
\usepackage{epstopdf, epsfig}
\usepackage{dcolumn}   
\usepackage{bm}        
\usepackage{verbatim} 
\usepackage{tikz}
\usepackage{overpic}
\usepackage{amsmath}
\usepackage{subfig}
\newcommand{\beq}{\begin{equation}}
\newcommand{\eeq}{\end{equation}}

\newcommand{\lb}{\left(}
\newcommand{\rb}{\right)}

\journalname{Journal of Engineering Mathematics}

\begin{document}

\title{Optimal control of diffuser shapes for non-uniform flow}

\author{G.P. Benham \and I.J. Hewitt \and C.P. Please \and P.A.D. Bird}


\institute{G.P. Benham \and I.J. Hewitt \and C.P. Please \at
              Mathematical Institute, University of Oxford, Andrew Wiles Building, Radcliffe Observatory Quarter, Woodstock Road, Oxford OX2 6GG United Kingdom \\
              \email{benham@maths.ox.ac.uk}           
           \and
           P.A.D. Bird \at
              VerdErg Renewable Energy Limited, 6 Old London Rd, Kingston upon Thames KT2 6QF, United Kingdom
}

\date{Received: date / Accepted: date}



\maketitle

\begin{abstract}
{A simplified model is used to identify the diffuser shape that maximises pressure recovery for several classes of non-uniform inflow.  Wide diffuser angles tend to accentuate non-uniform flow, causing poor pressure recovery, whilst shallow diffuser angles create enhanced wall drag, which is also detrimental to pressure recovery.  Optimal diffuser shapes strike a balance between these two effects, and the optimal shape depends on the structure of the non-uniform inflow.

Three classes of non-uniform inflow are considered, with the axial velocity varying across the width of the diffuser entrance.  The first case has inner and outer streams of different speeds, with a velocity jump between them that evolves into a shear layer downstream.  The second case is a limiting case when these streams are of similar speed.  The third case is a pure shear profile with linear velocity variation between the centre and outer edge of the diffuser.

We describe the evolution of the flow profile using a reduced mathematical model that has been previously tested against experiments and computational fluid dynamics (CFD) models.
The governing equations of this model form the dynamics of an optimal control problem where the control is the diffuser channel shape. A numerical optimisation approach is used to solve the optimal control problem and Pontryagin's Maximum Principle is used to find analytical solutions in the second and third cases. 
We show that some of the optimal diffuser shapes can be well approximated by piecewise linear sections. This suggests a low-dimensional parameterisation of the shapes, 
providing a structure in which more detailed and computationally expensive  turbulence models can be used to find optimal shapes for more realistic flow behaviour.}


\end{abstract}

\section{Introduction}

In this study, we consider a class of expanding channel flows in which the inflow is non-uniform. Expanding channels, known as diffusers, have the function of converting high-speed low-pressure flow to low-speed high-pressure flow. Diffusers have numerous applications, from turbines in aerospace and hydropower \cite{simone2012analysis,chamorro2013interaction,kang2014onset} to automotive design \cite{jones2003fluid}. There is a large literature on diffusers in the case where the inflow is uniform (see \cite{blevins1984applied}), but a limited literature for non-uniform inlet flows  \cite{benham2017turbulent}.

In the case where the inflow is uniform, diffusers are usually designed to be straight sided, and the expansion angle is critical to performance \cite{blevins1984applied}. The optimal angle strikes a balance between not being too shallow, since thin channels have larger wall drag, and not being too wide, since wide expansion angles result in boundary layer separation and poor consequent pressure recovery \cite{douglas1986fluid}. The optimum angle varies slightly, depending on the inflow boundary layer thickness, and whether the diffuser is two-dimensional or axisymmetric.

In the case where the inflow is non-uniform, we must account for the additional effect of the channel shape on the development of the non-uniform flow profile. An important feature in understanding non-uniform flows is the interplay between changes in the pressure and the kinetic energy flux factor, which is a normalised measure of how non-uniform a flow is. A decrease in kinetic energy flux factor corresponds to a more uniform flow, and a rise in pressure. Diffusers with wide angles have the tendency to accentuate non-uniform flows  and, in some extreme cases, create a jet-like outflow \cite{blevins1984applied}. In such cases, the outflow has a high kinetic energy flux and, hence, a low pressure recovery. On the other hand, diffusers with shallow angles have longer, narrower profiles, which create a lot of wall drag and consequently a larger drop in pressure. Optimal diffuser shapes, therefore, must strike a balance between mixing the flow in a narrow section and then widening the flow to decrease wall drag.

In this paper, we identify the optimal diffuser shape which satisfies these criteria. 
In contrast to diffusers with uniform inflow, where the channel shape is only restricted due to boundary layer separation, diffusers with non-uniform inflow have a shape which is also restricted due to the effect of accentuating the non-uniform flow. We find that in some cases, the optimum diffuser angle for non-uniform flow is smaller than typically used for diffusers with uniform inflow. {Furthermore, we show that, unlike for uniform flow, optimal diffuser shapes for non-uniform flow may contain an initial straight section that helps mix the flow before diffusing.} 
Therefore, from a design perspective, the effect of the inflow profile cannot be ignored.
In our analysis, we show how to optimise diffuser design based on the nature of the non-uniform inflow.

{We investigate three different classes of non-uniform inflow, with the axial velocity varying across the width of the diffuser entrance. The first case has inner and outer streams of different speeds, with a velocity jump between them that evolves into a shear layer downstream, and the shear layer eventually interacts with the channel walls.
The second case is the limit where the speeds of the streams are similar, creating a thin, slowly growing shear layer. 
In the third case, the inflow is a pure shear profile, with linear velocity variation between the centre and outer edge of the diffuser. These flow profiles are motivated by a low-head hydropower application, where the inner and outer streams are formed by a Venturi pipe which accelerates part of the flow in order to amplify the pressure drop across a turbine.}

For these non-uniform flows which we consider, the development of the flow profile, which is fundamental to pressure recovery, can be described using a simple model for turbulent shear layers in confining channels \cite{benham2017turbulent}. The model predictions have good agreement with CFD and experimental work for a range of channel shapes and Reynolds numbers. The model constitutes a differential-algebraic system of equations which governs the continuous dependance of the flow profile and pressure on the diffuser shape. It assumes that the flow is composed of uniform streams separated by a linear shear layer. {Wall drag is incorporated into the model with a friction factor, and the growth shear layers is modelled with a spreading parameter.}

Benham et al. \cite{benham2017turbulent} used this model to investigate pressure recovery of a simple class of diffuser shapes by exhaustively searching a restricted design space.  When we widen the parameter space and treat the diffuser shape as a continuous control, it is necessary to seek more
complex tools to solve the problem. In this paper, we use the model as the basis for numerical optimisation of the diffuser shape, where the governing equations form the optimisation constraints. Such problems, as well as PDE-constrained optimisation problems, often arise in the field of flow control. With the advancement of computational power, these problems have become more feasible to solve. There are many different approaches to solving such problems which are discussed by Gunzburger \cite{gunzburger2002perspectives}.


%

 
In our approach we exploit the fact that the model is one-dimensional and, upon discretisation, there are relatively few decision variables. This, in combination with the use of automatic differentiation to calculate gradients, allows us to use an interior point Newton method with relatively low computational effort \cite{nocedal2006numerical}. 
In certain limiting cases, we solve the optimal control problem analytically using Pontragin's maximum principle \cite{pontryagin1987mathematical} and  these analytical results aid interpretation of the results from the numerical optimisation. 
{We show that some of the optimal diffuser shapes look approximately like they are composed of piecewise linear sections. This motivates a low-dimensional parameterisation of the diffuser shapes, for which we use more detailed and computationally intensive CFD models to search for optima under more realistic flow behaviours. The two CFD models we use are 
a $k$-$\epsilon$ \cite{launder1974numerical} and a $k$-$\omega$ Shear Stress Transport (SST) \cite{menter1992improved} turbulence model. We find that the optimal diffuser shapes for both these CFD models are very similar to those found using our reduced model.}

Section \ref{model_opt} outlines the model for the non-uniform flow profiles we consider and sets up the optimal control problem, discussing the choice of objective, the constraints and the number of  parameters. Section \ref{numerical} outlines a numerical method for solving the optimal control problem and, using this method, we investigate optimal diffuser shapes in three different cases. In Section \ref{analytical} we find analytical solutions to the optimal control problem in the last two of these cases. In Section \ref{cfd_comp_sec}, we present some CFD calculations and compare them to the results of the optimisation. Section \ref{discuss} summarises the results of the paper and discusses the dependance of the optimal shapes on parameter choices.


\section{The model and optimal control problem\label{model_opt}}

\subsection{Modelling turbulent shear layers in confining channels
\label{model_model}}

{In this section we describe the flow scenarios which we consider and outline the simple model, previously presented by Benham et al.\cite{benham2017turbulent}, which we use to describe these flows. 
This model is based on integrated conservation of mass and momentum equations in a long and thin geometry, as well as Bernoulli's equation, which govern an idealised time-averaged flow profile. A friction factor is used to parameterise the effect of wall drag, whilst a spreading parameter models the growth of shear layers.

There are three different types of non-uniform channel flow we consider, all of which are symmetric about the channel centreline. In Fig. \ref{panel} we display each case, illustrating the axial velocity varying across the width of the diffuser. The first, which we call the \textit{Developing shear layer case}, is an inflow composed of inner and outer streams of different speeds, with a velocity jump in between. In this case, a shear layer forms between the streams and grows downstream, eventually interacting with the channel walls. We restrict our attention to situations where the inner stream is slower than the outer stream. In other situations where the outer stream is slower than the inner stream, there is a greater risk of boundary layer separation, since the slowest region of flow is next to the wall \cite{blevins1984applied}. Furthermore, it is well known that asymmetric flow instabilities, such as the Coanda effect \cite{tritton2012physical}, can occur in these situations, which we do not try to model here. 
The second case, called the \textit{Small shear limit}, is similar to the first case, except the inner and outer streams have near-identical velocities, such that the shear between the flows is small and the thin shear layer grows slowly. In the third case, called the \textit{Pure shear limit}, we consider a pure shear profile with linear velocity variation between the centre and outer edge of the diffuser.
This corresponds to the first case in the downstream limit, where the shear layer has reached across the entire channel. }

\begin{figure}
\vspace{0.5cm}
\centering
\centering
\begin{overpic}[width=0.3\textwidth]{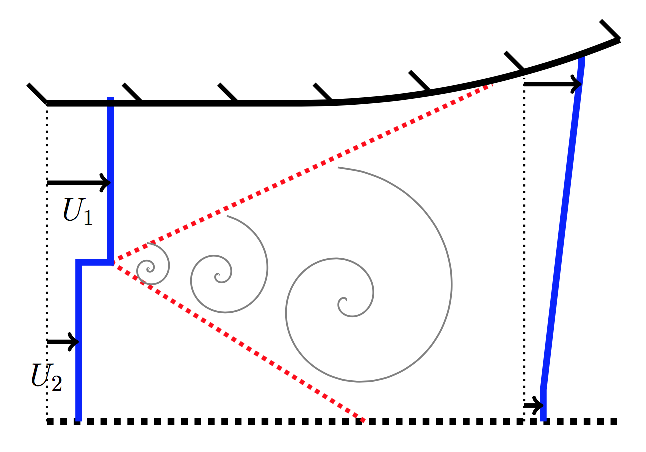}
\put(-10,70){(a) \textbf{Developing shear layer}}
\put(0,60){\tiny Wall $y=h(x)$}
\put(0,0){\tiny Centreline $y=0$}
\put(-10,30){\color{black}\vector(0,1){10}}
\put(-10,30){\color{black}\vector(1,0){10}}
\put(-17,35){\tiny $y$}
\put(-5,25){\tiny $x$}
\end{overpic}
\begin{overpic}[width=0.3\textwidth]{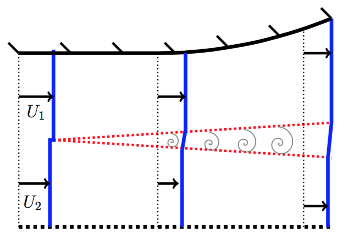}
\put(10,70){(b) \textbf{Small shear limit}}
\end{overpic}
\begin{overpic}[width=0.3\textwidth]{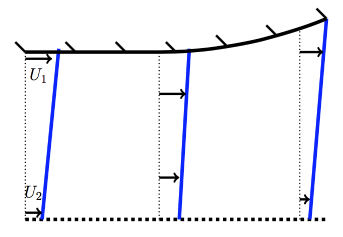}
\put(10,70){(c) \textbf{Pure shear limit}}
\end{overpic}
\caption{Schematic diagram of the different flow cases. a) \textit{Developing shear layer case}, where the inflow has inner and outer streams of different speeds, with a velocity jump between them that develops into a shear layer. The channel length is sufficiently large such that the growing shear layer reaches across the channel. b) \textit{Small shear limit} case, which is the limiting case where the speeds of the streams are similar, such that the thin, slowly growing shear layer never reaches across the channel. c) \textit{Pure shear limit}, where the velocity varies linearly between the centre and the outer wall of the diffuser. \label{panel}}
\end{figure}
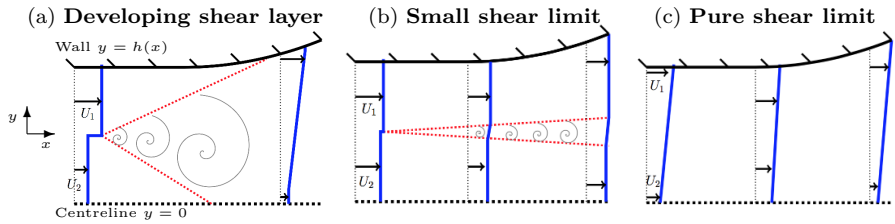

\begin{figure}
\centering
\begin{tikzpicture}[scale=0.8]
\draw [thick,gray] (3,2.5) arc [radius=0.2, start angle=45, end angle= 120];
\draw [thick,gray] (3.2,2.5) arc [radius=0.3, start angle=200, end angle= 320];
\draw [thick,gray] (4.0,2.5) arc [radius=0.4, start angle=45, end angle= 140];
\draw [thick,gray] (4.2,2.5) arc [radius=0.5, start angle=200, end angle= 350];
\draw [thick,gray] (6.0,2.5) arc [radius=1.0, start angle=45, end angle= 130];
\draw [thick,gray] (5.5,2.4) arc [radius=1.1, start angle=200, end angle= 360];
\draw [thick,gray] (9,2.2) arc [radius=1.1, start angle=0, end angle= 180];
\draw [thick,gray] (8.0,2.0) arc [radius=1.3, start angle=180, end angle= 380];
\draw[dashed,line width=3] (1,0) -- (10,0); 
\draw[red,dotted,line width=2] (2,5/2) -- (10,4); 
\draw[red,dotted,line width=2] (2,5/2) -- (10,0); 
\draw[blue, line width=3] (3/2, 0) -- (3/2,5/2) -- (2,5/2) -- (2,5.1) ; 
\draw[blue, line width=3] (8.8, 0) -- (8.8,0.5) -- (9.4,3.8) -- (9.4,6) ; 
\draw[->,line width=2] (1,15/4) -- (2,15/4); 
\draw[->,line width=2] (1,5/4) -- (3/2,5/4); 
\draw[->,line width=2] (8.5,0.25) -- (8.8,0.25); 
\draw[->,line width=2] (8.5,5) -- (9.4,5); 
\draw[dotted,line width=1] (1,0) -- (1,5); 
\draw[dotted,line width=1] (8.5,0) -- (8.5,52/9); 
\draw[->,line width=2]  (8,-0.8) --(7,5/2); 
\draw[->,line width=2] (4,6.1) -- (5,4); 
\draw[->,line width=2] (4,6.1) -- (4,1); 
\node at (4,6.5) {\large Plug flow};
\node at (9,-1.2) {\large Shear layer};
\draw[<->,line width=2] (6,3.4) -- (6,5);
\draw[<->,line width=2] (6,1.3) -- (6,3.2);
\draw[<->,line width=2] (6,0) -- (6,1.1);
\node at (6.4,4.25) {\large $h_1$};
\node at (6.4,2.2) {\large $\delta$};
\node at (6.4,0.5) {\large $h_2$};
\node at (1.5,3.3) {\large $U_1$};
\node at (1,0.7) {\large $U_2$};
\node at (10,5) {\large $U_1$};
\node at (9.8,0.4) {\large $U_2$};
\node at (11,3.5) {\large $\varepsilon_y=\frac{U_1-U_2}{\delta}$};
\draw[->,line width=0.5] (10.5,3.2) -- (9.5,2.5);
\draw[black,line width=3,-] (1,5) parabola bend (5,5) (10,6);
\draw[line width=2] (1,5) -- (0.7,5.3); 
\draw[line width=2] (2.5,5) -- (2.2,5.3); 
\draw[line width=2] (4,5) -- (3.7,5.3); 
\draw[line width=2] (5.5,5) -- (5.2,5.3); 
\draw[line width=2] (7,5.2) -- (6.7,5.5); 
\draw[line width=2] (8.5,5.5) -- (8.2,5.8); 
\draw[line width=2] (10,6) -- (9.7,6.3); 
\draw[->,line width=1] (-1.5,1.1) -- (-0.5,1.1);
\draw[->,line width=1] (-1.5,1.1) -- (-1.5,2.1);
\node at (-0.3,1.1) {\large $x$};
\node at (-1.8,1.6) {\large $y$};
\node at (1.5, 5.7) {\large Wall $y=h(x)$};
\node at (2,-0.6) {\large Centreline $y=0$};
\end{tikzpicture}
\caption{Schematic diagram of symmetric flow in a half channel. We model the flow as plug flow regions separated by a linear turbulent shear layer. The model governs a reduced number of variables $U_1$, $U_2$, $h_1$, $h_2$, $\delta$, $\varepsilon_y$ and $p$, which are all functions of $x$. The aspect ratio is exaggerated for illustration purposes. \label{tikz}}
\end{figure}
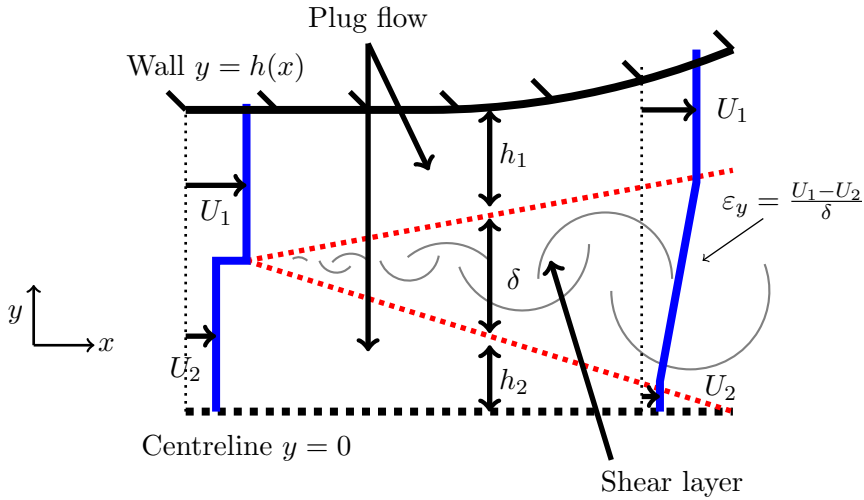

{The simple model, presented by Benham et al.   \cite{benham2017turbulent}, is used to describe the idealised flow profiles for each of the three cases. 
The first two cases share the same formulation, whilst the third case is slightly different. Thus, we start by describing the governing equations for the first and second cases.} Initially we consider two-dimensional flow in a half channel $0<y<h(x)$ and, later, we extend the model to axisymmetric channels (see Fig. \ref{panel}a, where we indicate our coordinate axes). The inflow for the first two cases is composed of a slower moving central stream with speed $U_2$ and a faster outer stream with speed $U_1$. A turbulent shear layer forms at the place where the parallel streams meet. We approximate the flow profile by decomposing it into two plug regions separated by a shear layer in which the velocity varies linearly between $U_1$ and $U_2$ (see Fig. \ref{tikz}). The approximate velocity profile is
\begin{equation}
{
u(x,y)=\begin{cases}
U_{2} (x)&: 0<y<h_2(x),\\
U_2 (x)+\varepsilon_y  (x) \lb y-h_2 (x)\rb &: h_2 (x)<y<h (x)-h_1 (x),\\
U_{1} (x) &: h (x)-h_1 (x)<y<h (x),
\end{cases}\label{piecewise}
}
\end{equation}
where $h_1$ and $h_2$ are the widths of the two plug regions, $\delta=h-h_1-h_2$ is the width of the shear layer, and $\varepsilon_y = (U_1-U_2)/\delta$ is the shear rate. {In the small shear limit, the plug flow speeds $U_{1}$ and $U_{2}$ are similar, such that the shear layer grows slowly. Whilst in the developing shear layer case, the shear layer may grow and interact with the channel walls, in the small shear limit, the channel is chosen to be sufficiently short that the slowly growing shear layer remains thin.} However, in both cases the shear rate decays with $x$ as the shear layer grows \cite{schlichting1960boundary}. We assume that the shear rate decays according to,
\beq
\frac{U_1+U_2}{2}\frac{d \varepsilon_y}{dx}=-S\varepsilon_y^2,\label{shear}
\eeq
where $S$ is a non-dimensional spreading parameter which must be determined from experiments or by comparison with CFD. Equation (\ref{shear}) can be derived from an entrainment argument (see Appendix in \cite{benham2017turbulent}), or by analogy with the growth of free shear layers. 
Assuming that the channel is long and thin, boundary layer theory \cite{schlichting1960boundary} indicates that, to good approximation, the pressure does not vary across the channel width $p=p(x)$. Averaged across the channel, conservation of mass and momentum equations are
\begin{align}
 \int_0^h \rho u\, dy = Q,\label{mass}\\
 \quad \frac{d}{d x}\lb  \int_0^h  \rho u^2\, dy \rb+h\frac{d p}{d x}=\tau_w,
\label{moment}
\end{align}
where $\rho$ is the density, $Q$ is the constant mass flux (per unit depth), and $\tau_w$ is the wall shear stress. 
We parameterise the
wall stress term with a friction factor $f$, such that $\tau_w=-1/8f\rho U_1^2$.
Finally, we ignore viscous dissipation in the plug flow regions, since it is small compared to that at the walls and in the shear layer. Hence, in the plug regions, we assume Bernoulli's equation holds \cite{batchelor2000introduction}.

\begin{figure}
\centering
\begin{tikzpicture}[scale=0.6]
\draw[dashed,line width=3] (1,0) -- (10,0); 
\draw[red,dotted,line width=2] (2,1.5) -- (8,5.3); 
\draw[red,dotted,line width=2] (2,1.5) -- (5,0) ; 
\draw[blue, line width=3] (3/2, 0) -- (3/2,1.5) -- (2,1.5) -- (2,3.3) ; 
\draw[blue, line width=3] (8.5, 0) -- (8.5,1.7)-- (9.4,5.6) -- (9.4,5.8) ; 
\draw[->,line width=2] (1,2) -- (2,2); 
\draw[->,line width=2] (1,0.5) -- (3/2,0.5); 
\draw[->,line width=2] (8.5,5.3) -- (9.4,5.3); 
\draw[dotted,line width=1] (1,0) -- (1,3); 
\draw[dotted,line width=1] (8.5,0) -- (8.5,5.4); 
\node at (1.5,2.6) {\large $U_1$};
\node at (1,1.1) {\large $U_2$};
\draw[black,line width=3,-] (1,3) -- (10,6);
\draw[line width=2] (1,3) -- (0.7,3.3); 
\draw[line width=2] (2.5,3.5) -- (2.2,3.8); 
\draw[line width=2] (4,4) -- (3.7,4.3); 
\draw[line width=2] (5.5,4.5) -- (5.2,4.8); 
\draw[line width=2] (7,5) -- (6.7,5.3); 
\draw[line width=2] (8.5,5.5) -- (8.2,5.8); 
\draw[line width=2] (10,6) -- (9.7,6.3); 
\draw[red,line width=2, dotted] (6,0) parabola bend(10,2) (10,2);
\node at (9.75,1) {\large $U_2=0$};
\draw [thick,gray] (2.8,1.9) arc [radius=0.7, start angle=180, end angle= 360];
\draw [thick,gray] (5.5,1.4) arc [radius=0.9, start angle=-20, end angle= 180];
\draw [thick,gray] (4.8,2.2) arc [radius=1.3, start angle=180, end angle= 400];
\end{tikzpicture}
\caption{Schematic diagram of stagnated flow in a diffuser, where the slower plug region decelerates and reaches zero velocity. This can occur if the inflow is sufficiently non-uniform, or if the diffuser angle is sufficiently large. The aspect ratio is exaggerated for illustration purposes.  \label{tikz2}}
\end{figure}
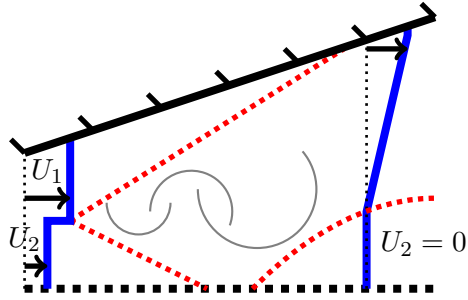

In certain cases, especially when the diffuser angle is wide, the speed of the slower plug region $U_2$ may decrease and reach zero. This has been observed in CFD simulations, which we display in Appendix \ref{appA}. In such cases there is a portion of recirculating flow in the central part of the diffuser.  We do not resolve the recirculation in these regions but since velocities are small, as observed in CFD, we treat the regions as stagnant zones with zero velocity (see Fig. \ref{tikz2}).

Bernoulli's equation {in each plug region holds along streamlines, ignoring transverse velocity components since they are small,} and is implemented in a complementarity format for convenience
\begin{align}
h_1\lb p-p(0)+\frac{1}{2}\rho (U_1^2- U_{1}(0)^2)\rb=0, &\quad \mathrm{and}\quad h_1\geq0,\label{bernoulli_stag}\\
U_2h_2\lb p-p(0)+\frac{1}{2}\rho (U_2^2- U_{2}(0)^2)\rb=0,&\quad \mathrm{and}\quad h_2\geq0, \quad U_2\geq0.\label{bernoulli_stag2}
\end{align}
The complementarity format of Eqs. (\ref{bernoulli_stag}) and (\ref{bernoulli_stag2}) ensures that when either of the plug regions disappears, or if the slower plug region stagnates, Bernoulli's equation ceases to hold in that region.
We find good comparison between our model predictions of the stagnant region and CFD calculations, which we discuss in Appendix \ref{appA}.

To summarise the first and second cases, the simple model describes the evolution of the non-uniform velocity profile $u(x,y)$, given by (\ref{piecewise}), and pressure $p(x)$ in a symmetric confining channel. Equations (\ref{shear}) - (\ref{bernoulli_stag2}) govern the variables $U_1$, $U_2$, $h_1$, $h_2$, $\delta$, $\varepsilon_y$ and $p$, which are all functions of $x$. These equations can be solved for all $x$ given inflow conditions at $x=0$. 
Since the shear layer forms at $x=0$, the inflow conditions for $\delta$ and $\varepsilon_y$ are $\delta(0)=0$ and $\varepsilon_y(0)=\infty$. Pressure is measured with reference to the value at the inlet so we can take $p(0)=0$ without loss of generality. All other inflow conditions form part of the set of parameters which we discuss in Section \ref{opt_con}.

{In the pure shear limit the plug regions are non-existent, such that $h_1=h_2=0$ and $\delta=h$. Then the velocity profile takes the form
\beq
u(x,y)=U_2(x)+\varepsilon_y(x)y.\label{pure_shear_prof}
\eeq
In this case the governing equations of the model reduce to (\ref{shear})-(\ref{moment}) and (\ref{pure_shear_prof}), which govern the variables $U_1$, $U_2$, $\varepsilon_y$ and $p$.}

We can extend the model to account for axisymmetric flows simply. For axisymmetric flow in a cylindrical channel $0\leq r\leq h$, we assume that that the velocity profile is identical to Eq. (\ref{piecewise}) in the first two cases, and (\ref{pure_shear_prof}) in the third case, except with $y$ replaced by $r$. In the axisymmetric version of the model, Eqs. (\ref{shear}) and (\ref{bernoulli_stag})-(\ref{bernoulli_stag2}) remain unchanged, but Eqs. (\ref{mass}) and (\ref{moment}) are altered to account for radial symmetry
\begin{align}
 2 \pi \int_0^h \rho u r\, dr = Q,\label{mass_r}\\
 \quad 2\pi \frac{d}{d x}\lb  \int_0^h  \rho u^2 r\, dr \rb+\pi h^2 \frac{d p}{d x}=2\pi h\tau_w.
\label{moment_r}
\end{align}
The results of the axisymmetric and two-dimensional cases are compared in  Section \ref{numerical}.

\subsection{Formulation of the optimal control problem \label{opt_con}}

{In this section we describe the optimal control problem by choosing an optimisation objective and formulating the control variables and contstraints.
Starting with the objective, we note that} diffuser performance can be measured in a number of different ways, for example using a 
pressure recovery coefficient or a loss coefficient \cite{blevins1984applied}. The pressure recovery coefficient $C_p$ is a measure of the pressure gain in the diffuser from inlet to outlet, relative to the kinetic energy flux at the inlet. The loss coefficient $K_l$ is a measure of the total energy lost from inlet to outlet, relative to the kinetic energy flux at the inlet. For our optimal control problem, we could choose either of these coefficients as the objective. Maximising $C_p$, for a given inflow, would produce the diffuser that converts the greatest amount of inflow kinetic energy into static pressure at the outflow. Minimising $K_l$, for a given inflow, would produce the diffuser with the maximum amount of energy at the outflow.

For this paper, we choose the pressure recovery coefficient as the objective. There are several ways to define the coefficient, but we shall use the so-called ``mass-averaged" pressure recovery \cite{filipenco1998effects}, which is defined as
\begin{equation}
C_p=\frac{\int_0^h u p \,dy |_{x=L} - \int_0^h u p \,dy |_{x=0}}{\int_0^h \frac{1}{2}\rho u^3 \,dy |_{x=0} },\label{C_p}
\end{equation}
for the two-dimensional case and
\begin{equation}
C_p=\frac{\int_0^h u p r\,dr |_{x=L} - \int_0^h u p r\,dr |_{x=0}}{\int_0^h \frac{1}{2}\rho u^3 r \,dr |_{x=0} },\label{C_p_axis}
\end{equation}
for the axisymmetric case.
The pressure recovery coefficient can take values $C_p\in[-\infty,1]$, where $C_p=1$ when all the kinetic energy of the inlet flow is converted into static pressure. For a given area ratio $h(L)/h(0)$ and inflow, there is a maximum possible pressure recovery $C_{p_I}\leq1$ \cite{blevins1984applied}. For uniform inviscid flow this ideal limit is $C_{p_I}=1-\lb h(0)/h(L)\rb^2$, but for non-uniform flow it is not known what the limit is. 

Now that we have chosen a suitable objective for the optimisation, we need to define a control. The diffuser shape is ultimately the control of the problem, but there are several different ways to formulate it. For example, we could use the shape function $h(x)$ as the control, or we could use its derivative, or even the second derivative. To aid our choice of control, we consider the regularity requirements of the final shape. If the minimum requirement is that the shape be continuous, it will be convenient to choose the derivative of $h$ as the control. If we also require smoothness (i.e. existence of the first derivative of $h$), then it will be convenient to choose the second derivative of $h$ as the control. However, if no such requirements exist, then it is satisfactory to use $h$ itself as the control. For this paper, we restrict ourselves to continuous but non-smooth shapes, and so we choose the shape derivative, or diffuser angle,
\beq
\alpha(x)=\frac{dh}{dx},
\eeq
as the control for optimising the diffuser shape. {In reality, sudden expansions and sharp corners, if severe enough, can cause flow separation which is detrimental to pressure recovery \cite{blevins1984applied}. Therefore, any such sharp corners must be rounded off with a suitable radius of curvature, upon construction.  However, we neglect this concern from our mathematical analysis.}
An additional possible control of the problem is the channel length $L$. For now, we consider this fixed, but later we discuss the possibility of including $L$ as a free parameter.

After defining both the objective and the control of the optimisation, we now discuss the constraints. The most obvious constraints on the variables are the governing equations and inflow conditions. In addition, we may also want some constraints on the outflow. As mentioned earlier, constraining $h(L)/h(0)$ gives us a fixed maximum value for the pressure recovery. If $h(L)/h(0)$ is unconstrained, then the pressure recovery will be maximised with $h(L)/h(0)=\infty$ \cite{blevins1984applied}. However, this is impractical for construction and, due to Bernoulli's equation, we see that pressure recovery decays rapidly with $h$ (like $\sim1/h^2$ for two-dimensional flows and like $\sim1/h^4$ for axisymmetric flows) so that a large majority of pressure is recovered for relatively small values of $h(L)/h(0)$. For example, if $h(L)/h(0)=3$  in uniform inviscid axisymmetric flow, the pressure recovery is $C_p\approx 0.99$. Therefore, for practical considerations, we constrain the channel width at the outflow
\beq
h(L)=h_L.\label{terminal}
\eeq
Another important constraint we need to consider is the boundedness of the control  $\alpha$. In particular, we note that for large values of the diffuser angle, boundary layers at the channel walls have the tendency to separate \cite{douglas1986fluid}. This phenomenon, which is often called `diffuser stall', is not something that we attempt to capture with our model. However, it is known that diffuser stall has a detrimental effect on pressure recovery (because the flow does not slow down). Considering this, we give the control $\alpha$ an upper bound corresponding to the smallest diffuser angle which causes stall. The first appreciable stall of a straight walled diffuser is at $\alpha\approx\tan 7^\circ$ for the two-dimensional case and $\alpha\approx\tan3.5^\circ$ for axisymmetric diffusers \cite{blevins1984applied}. 
Furthermore, due to engineering constraints, it might not always be possible to construct channel shapes which contract more than a certain angle. Therefore, a lower bound on the control may also be necessary.
If we denote the upper and lower bounds $\alpha_{max}$ and $\alpha_{min}$, respectively, then $\alpha$ satisfies the box constraints
\beq
\alpha_{min}\leq \alpha\leq \alpha_{max}.\label{alpha_bnd}
\eeq
It should be noted that, whilst Eq. (\ref{alpha_bnd}) applies, the optimal control might not necessarily attain these bounding values. In such cases, Eq. (\ref{alpha_bnd}) may be considered irrelevant.

To summarise the formulation of the optimal control problem, we seek to maximise the pressure recovery by manipulating the control $\alpha(x)$ within its bounds:
\beq
\max_{\alpha_{min}\leq \alpha(x)\leq \alpha_{max}} C_p,\label{opt_stat}
\eeq
with the constraints that Eqs. (\ref{shear})-(\ref{bernoulli_stag2}) hold, together with inlet conditions for all variables at $x=0$, and the end constraint (\ref{terminal}).

Before solving the optimal control problem, we note that there are several  parameters which affect the solution. We list these parameters in Table \ref{table} and discuss them in more detail in Section \ref{discuss}.
\begin{table}
\centering
\begin{tabular}{cc}
$U_2(0)/U_0$ & Velocity ratio\\
$h_2(0)/h_0$ & Plug width ratio\\
$h(L)/h_0$ & Expansion ratio\\
$L/h_0$ & Length ratio\\
$\alpha_{min},\, \alpha_{max}$ & Minimum/maximum angle\\
$S$ & Spreading parameter\\
$f$ &  Friction factor\\
\end{tabular}
\caption{List of the parameters of the optimal control problem. We treat the first $5$ of these parameters as problem-specific, whereas the last $2$ parameters are considered fixed. We also make use of the shorthand $h_0=h(0)$ and $U_0=U_1(0)$. \label{table}}
\end{table}

\section{Numerical optimisation \label{numerical}}

{This section, in which we solve the optimal control problem numerically for several different cases, is divided into subsections for clarity. Firstly, we describe our solution method for the numerical optimisation. Then, Section \ref{num_res} studies the solution in the developing shear layer case (see Fig. \ref{panel}a). In this case, we find that optimal diffuser shapes look approximately like they are composed of piecewise linear sections. This observation motivates us to introduce a low-dimensional parameterisation of the shapes that can be explored with contour plots, which is useful when comparing with CFD calculations later in Section \ref{cfd_comp_sec}. The small shear limit and the pure shear limit (see Fig. \ref{panel}b, c) are discussed in Sections \ref{small_shear_num} and \ref{pure_shear_num}, respectively. }

We solve the optimisation problem (\ref{opt_stat}) numerically by discretising space, introducing values of the variables $U_1$, $U_2$, $h_1$, $h_2$, $\delta$, $\varepsilon_y$ and $p$ at each spatial point, and treating each discretised value as a degree of freedom. We use an interior point Newton method \cite{nocedal2006numerical} (with the IpOpt library \cite{wachter2006implementation}) for non-linear constrained optimisation problems. Gradients are calculated using automatic differentiation in the the JuMP package \cite{dunning2017jump} of the Julia programming language \cite{bezanson2017julia}. The equality constraints we need to impose are Eqs. (\ref{shear})-(\ref{bernoulli_stag2}), inlet conditions and the terminal condition (\ref{terminal}). There are also the inequality constraints (\ref{alpha_bnd}) and those listed in the complementarity condition (\ref{bernoulli_stag})-(\ref{bernoulli_stag2}). As is often done, we impose the equality constraints using the quadratic penalty method \cite{nocedal2006numerical}, where we subtract their residual squared from the objective. For example, if our objective is to maximise the function $g(x)$ subject to the equality constraint $c(x)=0$, then we replace the objective with 
\beq
\max_{x} \quad g(x)-\mu c(x)^2,
\eeq
where $\mu$ is a penalty parameter. Our inequality constraints are simple box constraints. These are dealt with by the interior point method using logarithmic barrier functions. More details, including how to choose the penalty parameter $\mu$ and barrier functions, are discussed by Nocedal \& Wright \cite{nocedal2006numerical}.

It is not known whether this optimisation problem is convex so there may exist multiple solutions. In order to have confidence about the optimal solutions that are found, we use many different initial guesses to initialise the interior point method (although we have not yet found any multiple solutions).

The governing equations of the model consist of the algebraic equations, which are (\ref{mass}), (\ref{bernoulli_stag})-(\ref{bernoulli_stag2}) and inflow conditions, and the differential equations which are (\ref{shear}), and (\ref{moment}). We discretise space into $n$ points and impose the algebraic equations exactly at every point. The differential equations are imposed using a second order backward finite difference scheme. It should be noted that whilst the complementarity conditions enforce a switch in the governing equations and may produce non-smooth behaviour in the solution, the equations themselves are smooth and can therefore be differentiated. Computation time is fast, owing to the use of automatic differentiation to calculate gradients (as opposed to finite differencing, for example). With $8$ variables $U_1, U_2,h_1,h_2,\delta,\varepsilon_y,p,h$ and one control $\alpha$, the total number of degrees of freedom is $9n$. For a discretisation of $n=100$ grid points, and therefore $900$ degrees of freedom, computation time is of the order of less than $10$ seconds on a laptop computer.

\subsection{The developing shear layer case \label{num_res}}

Having discussed the optimisation routine, we use it to optimise channel shapes in several different cases, starting with the developing shear layer case. For plotting purposes, we maintain all variables in non-dimensional form with reference to typical length scales and velocity scales. We use the initial channel half-width as a typical length scale $h_0=h(0)$ and the speed of the faster plug region at the inlet as a typical velocity scale $U_0=U_1(0)$. 

\begin{figure}
\centering
\vspace{0.1cm}
\begin{overpic}[width=0.45\textwidth]{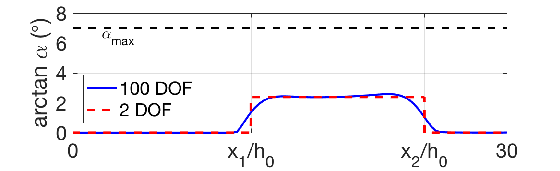}
\put(0,30){(a)}
\put(30,30){\bf Two-dimensional}
\end{overpic}
\begin{overpic}[width=0.45\textwidth]{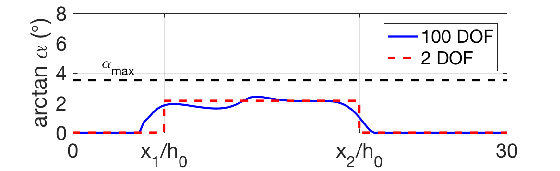}
\put(0,30){(b)}
\put(32,30){\bf Axisymmetric}
\end{overpic}\\
\vspace{0.4cm}
\begin{overpic}[width=0.45\textwidth]{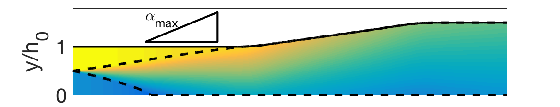}
\put(0,20){(c)}
\put(25,20){\bf Velocity (100 DOF)}
\end{overpic}
\begin{overpic}[width=0.45\textwidth]{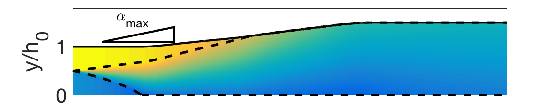}
\put(0,20){(d)}
\put(25,20){\bf Velocity (100 DOF)}
\end{overpic}\\
\vspace{0.5cm}
\begin{overpic}[width=0.45\textwidth]{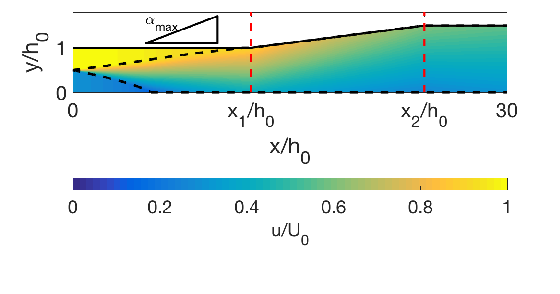}
\put(0,50){(e)}
\put(28,52){\bf Velocity (2 DOF)}
\end{overpic}
\begin{overpic}[width=0.45\textwidth]{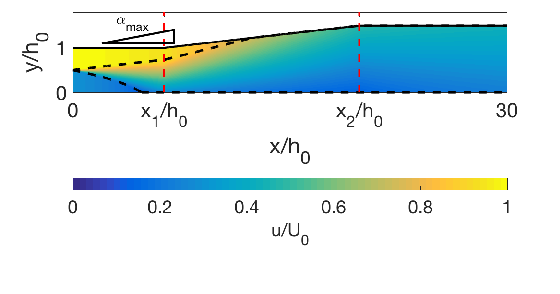}
\put(0,50){(f)}
\put(28,52){\bf Velocity (2 DOF)}
\end{overpic}\\
\begin{overpic}[width=0.45\textwidth]{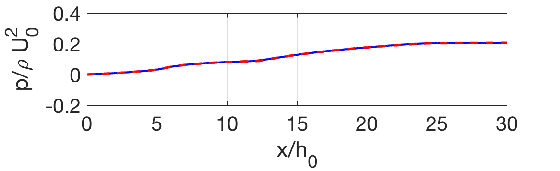}
\put(0,33){(g)}
\put(40,30){\bf Pressure}
\end{overpic}
\begin{overpic}[width=0.45\textwidth]{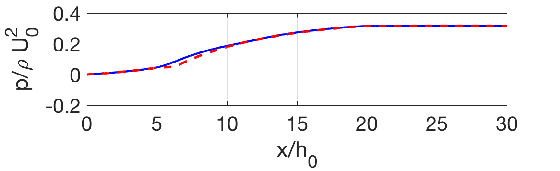}
\put(0,33){(h)}
\put(40,30){\bf Pressure}
\end{overpic}
\caption{Optimal diffuser shape for the developing shear layer case, in two dimensions (a, c, e, g) and in the axisymmetric case (b, d, f, h).  The optimal shapes found using $100$ degrees of freedom ($100$ DOF) (c, d) can be well approximated by three constant-angle sections with divisions at $x=x_1$ and $x=x_2$ ($2$ DOF), as shown by comparisons of diffuser angle (a, b), velocity (e, f) and pressure (g, h).  The parameter values used are given in the main text. \label{results1}}
\end{figure}
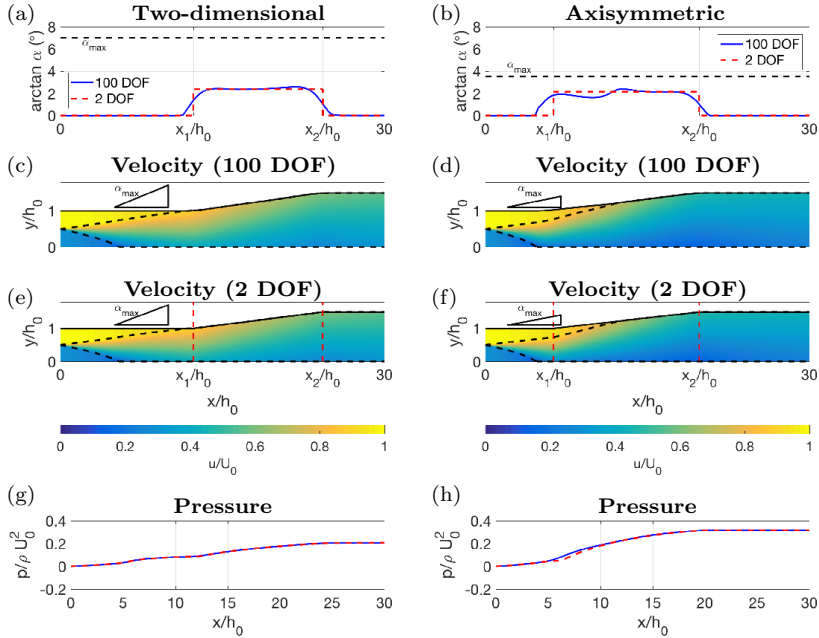

In the developing shear layer case, we look at two-dimensional flow and choose parameter values $U_2(0)/U_0=0.3$, $h_2(0)/h_0=0.5$, $h(L)/h_0=1.5$, $L/h_0=30$, $\alpha_{min}=0^\circ$ and $\alpha_{max}=\tan7^\circ$. The other parameters are taken as $S=0.11$ {(which we determine from comparison with CFD in Section \ref{cfd_comp_sec})} and $f=0.01$ (which corresponds to a Reynolds number of $Re=10^6$ and hydraulically smooth walls \cite{blasius1913ahnlichkeitsgesetz,mckeon2005new}). The number of grid points for discretising the  simple model is $n=100$. Simultaneously, we also investigate axisymmetric flow with the same parameter values, except with $\alpha_{max}=\tan3.5^\circ$. We plot the optimal diffuser angle for the two-dimensional case in Fig. \ref{results1}a, the corresponding optimal diffuser shape and velocity colour map in Fig. \ref{results1}c, and pressure plot in Fig. \ref{results1}g. The axisymmetric case is plotted in Fig. \ref{results1}b, d, h.

{In both cases, we observe that the optimal shape looks approximately like a piecewise linear function}, which is divided into a straight part, followed by a widening part, followed by another straight part. In the two-dimensional case, the length of the first straight section aligns with the length it takes for the shear layer to spread completely across the channel. This suggests that mixing the flow to a more uniform profile is advantageous for the widening part to perform well. This is as expected because, as mentioned earlier, diffusers tend to accentuate non-uniform flow, producing an outflow with large kinetic energy flux (and therefore a low pressure recovery). However, as discussed earlier, long thin channels cause large loss in pressure due to wall drag. Therefore, the optimal shape must have a straight section which is sufficiently long that the shear layer reaches across the channel, making the flow more well mixed, but no longer than that because of wall drag.

Interestingly, the widening part of the channel widens at a shallower angle than the maximum value (around $\tan2.3 ^\circ$ compared to $\tan7^\circ$). So the upper bound on $\alpha$ is not needed in this case. This behaviour is unexpected since diffusers are usually designed with a widening angle as close to $\tan7^\circ$ as possible, regardless of the inflow. These results suggest that there is an optimum widening angle which is determined by the non-uniform inflow, rather than the risk of boundary layer separation.

\begin{figure}
\centering
\vspace{0.5cm}
\begin{overpic}[width=0.45\textwidth]{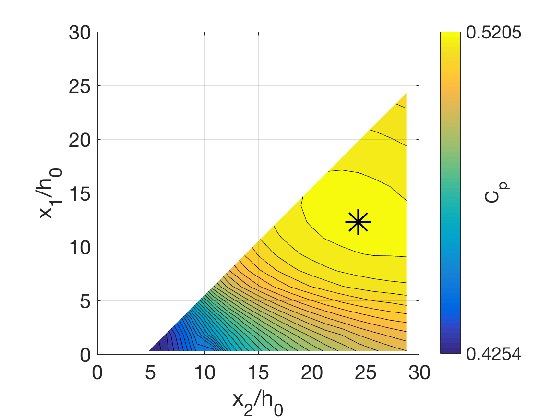}
\put(30,72){\bf Two-dimensional}
\put(0,65){(a)}
\end{overpic}
\begin{overpic}[width=0.45\textwidth]{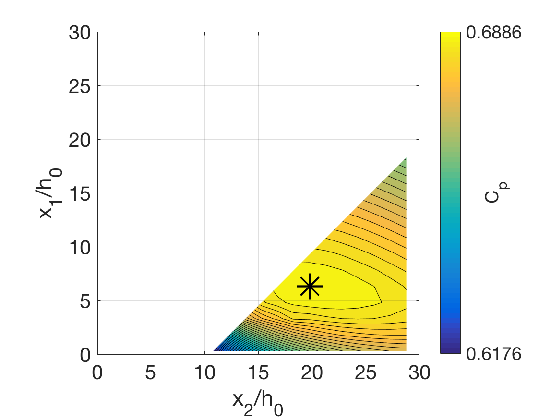}
\put(32,72){\bf Axisymmetric}
\put(36,32){\color{black}\vector(1,-2){9}}
\put(22,33){ Stagnation}
\put(0,65){(b)}
\end{overpic}
\caption{
 Contour plots of pressure recovery $C_p$ (\ref{C_p}), (\ref{C_p_axis}), using the low-dimensional parameterisation of the diffuser shapes, as shown in Fig. \ref{results1}a, b, for all permissible values of $x_1$ and $x_2$, such that $\alpha\leq \alpha_{max}$. (a) Two-dimensional case with $\alpha_{max}=\tan 7^\circ$.  (b) Axisymmetric case with $\alpha_{max}=\tan 3.5^\circ$, indicating where the inner stream stagnates. The widening middle section has constant angle given by Eq. (\ref{wide_angle}). \label{results2}}
\end{figure}
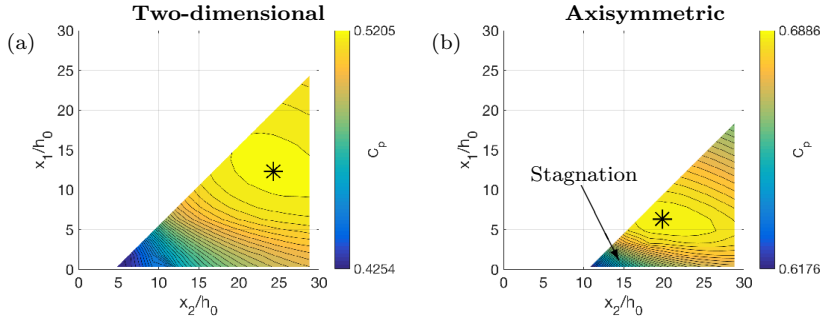

Since we observe that the optimal shape in Fig. \ref{results1}a, b, c, d looks approximately piecewise linear with three sections, we also try restricting the control $\alpha$ in this way to see if we can attain a near optimal solution with a piecewise linear shape. We parameterise $\alpha$ by splitting it into three parts: a straight part with $\alpha=0$ for $0\leq x <x_1$; a widening part with constant $\alpha>0$ for $x_1\leq x <x_2$, and a final straight part with $\alpha=0$ for $x_2\leq x\leq L$. We treat $x_1$ and $x_2$ as control parameters and the value of $\alpha$ in the middle section is determined by the condition
\beq
\alpha=\frac{h_L-h_0}{x_1-x_2}\label{wide_angle}.
\eeq
We optimise pressure recovery, using the same algorithm as before, but with $\alpha$ having only $2$ degrees of freedom (DOF), the two parameters $x_1$ and $x_2$, instead of $100$ DOF. 
We plot the optimal diffuser angle in Fig. \ref{results1}a, which is nearly identical to that obtained with $100$ DOF. Moreover, the velocity colour map and pressure plot displayed in Fig. \ref{results1}e, g both show a very close match. The pressure recovery coefficient for $2$ DOF is $C_p=0.5205$, which is the same as for $100$ DOF (up to $4$ decimal places), suggesting that piecewise linear diffuser shapes are a very good approximation in this case.
In Fig. \ref{results2}a we display a contour plot of $C_p$ for all possible values of $x_1$ and $x_2$ (we cut out part of the contour plot corresponding to $\alpha>\tan7^\circ$). This indicates that there is a clear unique optimum at $x_1/h_0=12.3$ and $x_2/h_0=24.3$.  {Note that an unoptimised shape, say with $x_1=0$ and $x_2=5$, gives a value of $C_p=0.4254$, which is $22\%$ worse than the optimal shape.}

For the axisymmetric case we find that the optimal shape has a similar structure and can also be well approximated by parameterisation with $x_1$ and $x_2$. We find that the optimal value of $x_1/h_0=6.3$ is a little bit shorter than the two-dimensional case. In fact, we can see in Fig. \ref{results1}d, f that the shear layer has not reached all the way across the channel by $x=x_1$. Instead, $x_1$ corresponds to the point where the shear layer reaches the centre of the channel. It reaches the outer wall of the channel slightly further downstream, during the widening section. This could be explained by the fact that pressure gradients due to wall drag are stronger (per unit flux) in the axisymmetric case (e.g. Poiseuille flow \cite{batchelor2000introduction}), and therefore the optimal shape cannot afford a longer section of straight, narrow channel. Figure \ref{results1}b, d, f, h shows a close comparison between the diffuser angle, velocity and pressure in the $2$ DOF case and the $100$ DOF case. The contour plot in Fig. \ref{results2}b shows the optimum parameter values $x_1/h_0=6.3$ and $x_2/h_0=19.8$. The pressure recovery coefficient for $2$ DOF is $C_p=0.6886$ compared to $C_p=0.7018$ for $100$ DOF, suggesting that axisymmetric piecewise linear diffuser shapes are also a good approximation, but slightly less so than in the two-dimensional case.


{The contour plots in Fig. \ref{results2} have a steep gradient for small $x_1/h_0$, indicating that diffuser performance is poor in this case.  This corresponds to situations in which the inner stream stagnates.  The phenomenon of stagnation is particularly an issue when the inner stream is slow, and this analysis shows that the way to avoid such poor performance is to have a longer straight section in which the inner stream is accelerated before diffusing. }

\begin{figure}
\centering
\vspace{0.5cm}
\begin{overpic}[width=0.45\textwidth]{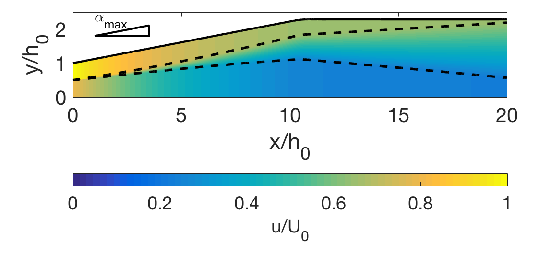}
\put(0,45){(a)}
\put(30,49){\bf Small shear limit}
\end{overpic}
\begin{overpic}[width=0.45\textwidth]{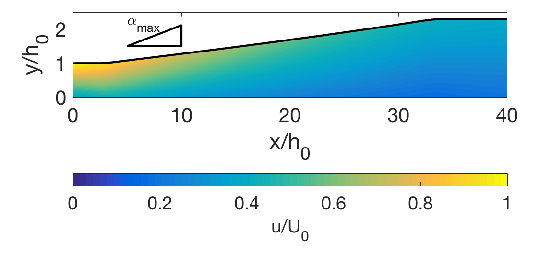}
\put(0,45){(b)}
\put(30,49){\bf Pure shear limit}
\end{overpic}\\
\begin{overpic}[width=0.45\textwidth]{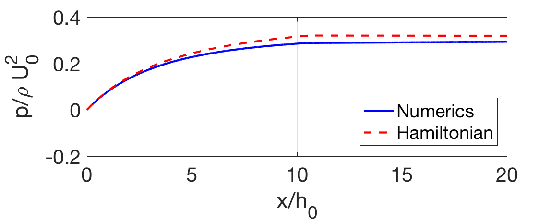}
\put(0,40){(c)}
\end{overpic}
\begin{overpic}[width=0.45\textwidth]{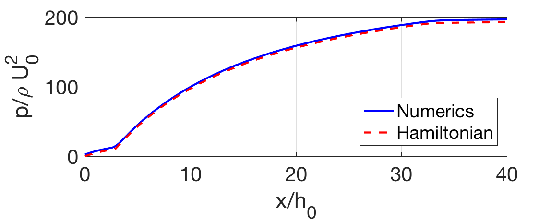}
\put(0,40){(d)}
\end{overpic}\\
\begin{overpic}[width=0.45\textwidth]{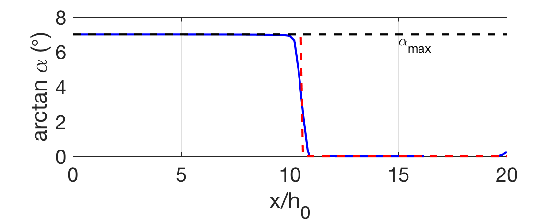}
\put(0,40){(e)}
\end{overpic}
\begin{overpic}[width=0.45\textwidth]{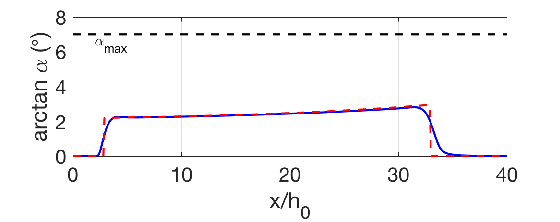}
\put(0,40){(f)}
\end{overpic}
\caption{Optimal diffuser shapes, velocity colour maps, pressure plots and diffuser angle plots for the small shear limit and the pure shear limit. (a, c, e) Two-dimensional flow where the plug regions have similar speeds $U_2(0)/U_0=0.8$ and the channel length $L/h_0=20$ is sufficiently short such that the shear layer never reaches across the channel. (b, d, f) Two-dimensional flow where the shear layer has already reached across the channel at the inflow so that there are no plug regions. The inflow velocity ratio is $U_2(0)/U_0=0.35$.  \label{results3}}
\end{figure}

\subsection{Small shear limit\label{small_shear_num}}


For the small shear limit, we consider two-dimensional flow and choose parameter values $U_2(0)/U_0=0.8$, $h_2(0)/h_0=0.5$, $h(L)/h_0=2.3$, $L/h_0=20$, $\alpha_{min}=0$ and $\alpha_{max}=\tan7^\circ$. The other parameters $S$ and $f$ are taken at the same values as the previous cases. We display the optimal shape, velocity colour map and pressure plot in Fig. \ref{results3}a, c, e. In this case, the optimal shape widens at the maximum angle $\alpha_{max}$ until it reaches the exit width $h(L)$ and then stays straight. Since the shear is small and the flow is almost uniform, there is no risk of accentuating the flow profile drastically. Therefore, wide angles initially are not penalised much, allowing the control to take its maximum value. This design is typically what is built in the diffuser industry for uniform inflow \cite{blevins1984applied}, where the maximum diffuser angle is set by the limit where boundary layer separation occurs. The optimal control only takes its extremal values, which is sometimes referred to as ``bang-bang control". In Section \ref{small_shear} we discuss analytical results for this limiting case and prove that the control must be bang-bang.

\subsection{Pure shear limit\label{pure_shear_num}}


For the pure shear limit, we consider two-dimensional flow and choose parameter values $U_2(0)/U_0=0.35$, $h_L/h_0=2.3$, $L/h_0=40$, $\alpha_{min}=0^\circ$ and $\alpha_{max}=\tan7^\circ$. Parameters $S$ and $f$ are taken as the same as before. The optimal channel, velocity colour map, pressure and control are displayed in Fig. \ref{results3}b, d, f. The optimal shape is similar to those in Fig. \ref{results1}, with a natural decomposition into two straight sections separated by a widening section. The widening section has an angle that increases from $\alpha\approx \tan2.5^\circ$ to $\alpha\approx\tan3^\circ$ and is nowhere greater than $\tan7^\circ$, showing that the upper bound $\alpha$ was not needed in this case. The optimal shape, like in Fig. \ref{results1}, exhibits the balance between the necessity of a straight section that is long enough to allow some mixing, but not too long that wall drag dominates. In this case, which does not involve any of the switching  behaviour that occurs when plug regions reach the wall, we derive some analytical results (discussed in Section \ref{pure_shear}) which support and help interpret the numerical optimisation. In particular, we investigate the nature of the optimal widening angle which lies in the interval $(\alpha_{min},\alpha_{max})$. This is of great interest because it indicates that the optimal design is unaffected by the conventional widening angle limit that exists due to boundary layer separation.


\section{Analytical results \label{analytical}}

The numerical optimisation routine outlined in Section \ref{numerical} can be applied to find optimal shapes for any choice of the parameters listed in Table \ref{table}. We have seen several examples of these in Figs. \ref{results1} and \ref{results3}. In this section we show that in the two limiting cases displayed in Fig. \ref{results3}, the small shear limit and the pure shear limit, it is possible to make some analytical progress which aids our understanding and interpretation of the optimal control. Furthermore, the results discussed in this section include simple relationships that may be of instructive use for the purpose of diffuser design in industry. 
In both cases we derive a reduced set of equations describing the dynamics, that is amenable to optimal control analysis using Pontryagin's maximum principle \cite{pontryagin1987mathematical}.

\subsection{Small shear limit \label{small_shear}}

To start with, we consider a two-dimensional diffuser where the inflow, given by (\ref{piecewise}), is almost uniform, producing a thin, slowly growing shear layer between the plug flow regions. The channel is sufficiently short that the shear layer never reaches across the channel (see Fig. \ref{panel}b). The speed of the plug regions differs by a small amount 
\beq
U_0-U_2(0)=\epsilon V,
\eeq
where $\epsilon\ll1$. If the plug regions always exist, with positive width $h_1>0$, $h_2>0$, and we assume that the slower flow never stagnates $U_2>0$, then we need not consider the complementarity format for Bernoulli's equation. Therefore, we replace Eqs. (\ref{bernoulli_stag}) and (\ref{bernoulli_stag2}) with
\beq
p+\frac{1}{2}\rho U_i^2=\frac{1}{2}\rho U_{i}(0)^2, \quad \mathrm{for}\quad  i=1,2.\label{bernoulli_no_switch}
\eeq
We consider the distinguished limit where the friction factor $f$ is small such that $f=\epsilon S F $, where $F=O(1)$. For a Reynolds number of $Re=10^6$ and hydraulically smooth walls, the friction factor is $f=0.01$. Therefore, if $\epsilon = 0.1$ and $S=0.11$, then $F=0.91$. Choosing these parameter values and setting $V/U_0=2$, we achieve the small shear limit example in Fig. \ref{results3}a, c, e.
We expand variables in powers of the small parameter $\epsilon$,
\begin{align}
U_1&=U_{1_0}+\epsilon \hat{U}_1+\ldots,\label{ass1}\\
U_2&=U_{2_0}+\epsilon \hat{U}_2+\ldots,\\
h_1&=h_{1_0}+\epsilon \hat{h}_1+\ldots,\\
h_2&=h_{2_0}+\epsilon \hat{h}_2+\ldots,\\
p&=p_0+\epsilon \hat{p}+\ldots,\\
\delta&=\delta_0+\epsilon \hat{\delta}+\ldots,\\
\varepsilon_y&=\varepsilon_{y_0}+\epsilon\hat{\varepsilon}_y+\ldots.\label{ass7}
\end{align}
In the limit $\epsilon\rightarrow 0$, Eqs. (\ref{shear})-(\ref{moment}) and (\ref{bernoulli_no_switch}) are satisfied by
\begin{align}
U_{1_0}&=U_{2_0}=\frac{U_0h_0}{h},\\
h_{2_0}&=h-h_{1_0},\\
p_0&=\frac{1}{2}\rho U_0^2\lb1-\frac{h_0^2}{h^2}\rb,\\
\delta_0&=0,\\
\varepsilon_{y_0}&=\frac{h_0U_0}{S\int_0^x h(\hat{x})\,d\hat{x}}.
\end{align}
The function $h_{1_0}$, which represents the location of the centre of the shear layer to leading order, is determined at order $O(\epsilon)$. Bernoulli's equation (\ref{bernoulli_no_switch}) for each plug region, at order $O(\epsilon)$, is
\begin{align}
\hat{p}+\rho \hat{U}_1 U_{1_0} &=0,\label{bernoulli_1}\\
\hat{p}+\rho \hat{U}_2 U_{1_0} &=-\rho U_0 V.\label{bernoulli_2}
\end{align}
From the relationship $h_1+h_2+\delta=h$ at order $O(\epsilon)$, we find that
\beq
\hat{h}_1 +\hat{h}_2 + \hat{\delta}=0.\label{regions}
\eeq
Thus, the conservation of mass equation (\ref{mass}) is
\beq
\hat{U}_1h_{1_0} + \hat{U}_2 (h - h_{1_0}) = -V h_2(0),\label{mass_eps}
\eeq
and the momentum equation (\ref{moment}) is
\beq
h\frac{d\hat{p}}{dx}+\rho\frac{d}{dx}\lb 2 U_{1_0}\hat{U}_1 h_{1_0} +2 U_{1_0}\hat{U}_2 (h-h_{1_0})  \rb=-\frac{1}{8}SF\rho U_{1_0}^2.\label{mom_eps}
\eeq
Thus, using Eqs. (\ref{bernoulli_1}) - (\ref{regions}), we can simplify Eq. (\ref{mom_eps}) to an equation purely involving $h_{1_0}$ and $h$
\beq
V h^2 \frac{dh_{1_0}}{dx} -V h h_{1_0} \frac{dh}{dx} =-\frac{1}{8}F Sh_0^2 U_0.\label{H_1_eq}
\eeq
Equation (\ref{H_1_eq}) has solution
\beq
h_{1_0}=h\lb \frac{h_1(0)}{h_0} -\int_0^x \frac{SFh_0^2U_0}{8V h(\hat{x})^3}\,d\hat{x} \rb.
\eeq
Combining (\ref{mass_eps}) and (\ref{mom_eps}) we obtain a differential equation for the pressure correction $\hat{p}$ in terms of the channel shape and its derivative $\alpha=h'(x)$,
\beq
\frac{d\hat{p}}{dx}=-\frac{\rho  U_0^2 h_0^2 }{h^{3}} \lb   \frac{2 V}{U_0} \lb1- \frac{h_1(0)}{h_0}\rb \alpha + \frac{S F}{8} \rb .\label{p_eqn}
\eeq
We now solve the optimal control problem outlined in Section \ref{opt_con}. Since the inflow conditions are fixed and we take $p(0)=0$, maximising $C_p$ is equivalent to maximising pressure at the outlet $p(L)$. Furthermore, the constraint equations have been reduced to (\ref{p_eqn}). Therefore the optimal control problem, including terms up to and including order $O(\epsilon)$, and written as a system of first order differential equations, is as follows:
\beq
\max_{\alpha_{min} \leq \alpha(x)\leq \alpha_{max}}\quad \Phi: = p(L),
\eeq
such that
\begin{align}
\frac{dp}{dx}&=\frac{\rho  U_0^2 h_0^2 }{h^{3}} \lb \lb 1- \epsilon\frac{2 V}{U_0} \lb1- \frac{h_1(0)}{h_0}\rb\rb \alpha -\epsilon \frac{S F}{8}  \rb ,\label{gov1}\\
\frac{dh}{dx}&=\alpha,\label{gov2}\\
h(0)&=h_0,\label{govbc1}\\
p(0)&=0,\label{govbc2}\\
h(L)&=h_L.\label{govbc3}
\end{align} 
We now solve this reduced problem using Pontryagin's maximum principle \cite{pontryagin1987mathematical}. The Hamiltonian for this system is
\beq
H=\lambda_p \frac{dp}{dx}+\lambda_h \frac{dh}{dx},\label{ham}
\eeq
where $\lambda_p$ and $\lambda_h$ are the adjoint variables which satisfy the adjoint equations
\begin{align}
\frac{d\lambda_p}{dx}&=-\frac{\partial H}{\partial p},\label{ad1}\\
\frac{d\lambda_h}{dx}&=-\frac{\partial H}{\partial h}\label{ad2}.
\end{align} 
According to Pontryagin's maximum principle, variables which appear in the objective function evaluated at $x=L$ must have natural boundary conditions which apply to their corresponding adjoint variables \cite{pontryagin1987mathematical}. Since the objective function only depends on pressure at the outlet $\Phi=p(L)$, we have the natural boundary condition
\beq
\lambda_p(L)=\frac{\partial \Phi}{\partial p}=1.\label{lambda_p}
\eeq
Considering Eq. (\ref{lambda_p}) and the fact that there is no dependance of the Hamiltonian (\ref{ham}) on the pressure $p$, Eq. (\ref{ad1}) tells us that $\lambda_p=1$ for all values of $x$. There is no natural boundary condition for $\lambda_h$ since we are enforcing a condition on $h$ at the outlet $x=L$. The last condition from Pontryagin's maximum principle is the optimality condition, which usually takes the form $\partial H/\partial \alpha=0$. However, following Pitcher \cite{pitcher} and McDanell \& Powers \cite{mcdanell1971necessary}, since the Hamiltonian is linear in the control $\alpha$, the optimality condition takes the form
\beq
\alpha(x)=\begin{cases}
    \alpha_{max},  & \text{if } H_\alpha>0, \\
    \in [\alpha_{min},\alpha_{max}],  & \text{if } H_\alpha=0, \\
    \alpha_{min},  & \text{if } H_\alpha<0, \\
\end{cases}\label{ham_cases}
\eeq
where we have introduced the shorthand notation $H_\alpha=\partial H/\partial \alpha$. If $H_\alpha$ is only zero at single values of $x$, the control is said to be ``bang-bang''. If the Hamiltonian satisfies $H_\alpha=0$ and $dH_\alpha/dx=0$ for a finite interval, then the control is said to have a ``singular arc''. However, upon close inspection, we see that
\beq
\frac{dH_\alpha}{dx}=-\frac{3 \epsilon \rho S F   U_0^2 h_0^2 }{8h^{4}},
\eeq
which is negative for all values of $x$. Hence, it is impossible for singular arcs to exist in this case. Therefore the control is bang-bang with
\beq
\alpha(x)=\begin{cases}
   \alpha_{max},  & \mathrm{for} \quad x\in[0,\gamma], \\
    \alpha_{min},   & \mathrm{for} \quad x\in[\gamma,L], \\
\end{cases}\label{cases}
\eeq
where the switching point $\gamma$ is given by
\beq
\gamma=\frac{h_L-h_0-\alpha_{min} L}{\alpha_{max}  - \alpha_{min} }\label{switch}.
\eeq

In Fig. \ref{results3}c, e we plot the solution to the optimal control problem found using the Hamiltonian approach on top of the solution found using the numerical optimisation routine outlined in Section \ref{numerical}. It is clear that the numerical optimisation routine has correctly found the bang-bang control which we have derived here, with $\gamma/h_0=10.59$ (the small discrepancy is probably due to the finite value of $\epsilon$).

It should be noted that the adjoint variable $\lambda_h$ is only solved for up to a constant of integration $C$ (from integrating Eq. (\ref{ad2})) since it has no boundary condition. Instead, $C$ is determined by the condition that
\beq
H_\alpha(\gamma)=0. \label{halphagamma}
\eeq 
However, in the case where we also allow the channel length $L$ to be a control as well as $\alpha$, according to Pontryagin's maximum principle \cite{pitcher,pontryagin1987mathematical}, we have the additional constraint on the Hamiltonian at the final point
\beq
H(L)=0.\label{term_con}
\eeq
It is straightforward to show that (\ref{halphagamma}) and (\ref{term_con}) are inconsistent unless $\gamma=0$ or $L$.
For the case in Fig. \ref{results3}a, c, e, it is clear that $\gamma=0$ is impossible, so we conclude that the optimal diffuser length is $L=\gamma$. Therefore, including $L$ as a control and taking $\alpha_{min}=0^\circ$, the optimal diffuser shape for the small shear limit is one which expands at the maximum angle until $h$ reaches $h_L$, at which point the channel terminates.


\subsection{Pure shear limit \label{pure_shear}}

The next limiting case we investigate is the pure shear limit, in which the shear layer has already reached across the channel at the inflow, such that there are no plug regions (see Fig. \ref{panel}c). The velocity profile is given by Eq. (\ref{pure_shear_prof}).
For this velocity profile, conservation of mass and momentum equations (\ref{mass}), (\ref{moment}) reduce to
\begin{align}
\frac{h}{2}(U_1+U_2)&=Q\label{pure_mass},\\
h\frac{dp}{dx}+\frac{1}{3}\rho\frac{d}{dx}\lb h\lb U_1^2+U_1U_2+U_2^2\rb \rb&=-\frac{1}{8}\rho f U_1^2.\label{pure_moment}
\end{align}
{We now solve the optimal control problem outlined in Section \ref{opt_con}. As in Section \ref{small_shear} we maximise pressure at the outlet $p(L)$. Furthermore, it is convenient to introduce a new variable, the scaled velocity difference between maximum and minimum velocities $\tilde{U}=(U_1-U_2)/(U_1+U_2)$. Using (\ref{pure_mass}) and this new variable we can simplify Eqs. (\ref{shear}) and (\ref{pure_moment}), which are the reduced system of constraint equations.}
Therefore, the optimal control problem is as follows:
\beq
\max_{\alpha_{min} \leq \alpha(x) \leq \alpha_{max}} \quad \Phi:=p(L),
\eeq
such that
\begin{align}
\frac{d\tilde{U}}{dx}&=\frac{2\tilde{U}(\alpha-S\tilde{U})}{h},\label{dyn1}\\
  \frac{dp}{dx}&=\frac{\rho Q^2\lb 32 S \tilde{U}^3 -3 f  (1 +  \tilde{U})^2 + 24 \alpha (1 - \tilde{U}^2)\rb}{24 h^3},\\
\frac{dh}{dx}&=\alpha,\label{dyn3}\\
\tilde{U}(0)&=\tilde{U}_0,\\
p(0)&=0,\\
h(0)&=h_0,\\
h(L)&=h_L,
\end{align} 
where $\tilde{U}_0=(1-U_2(0)/U_0)/(1+U_2(0)/U_0)$. Similarly to Section \ref{small_shear}, the Hamiltonian for the system is constructed as
\beq
H=\lambda_{\tilde{U}} \frac{d\tilde{U}}{dx}+\lambda_p \frac{dp}{dx}+\lambda_h \frac{dh}{dx},\label{pure_ham}
\eeq
which is linear in the control $\alpha$. The adjoint equations are
\begin{align}
\frac{d\lambda_{\tilde{U}}}{dx}&=-\frac{\partial H}{\partial \lambda_{\tilde{U}}},\label{pure_ad1}\\
\frac{d\lambda_p}{dx}&=-\frac{\partial H}{\partial p},\label{pure_ad2}\\
\frac{d\lambda_h}{dx}&=-\frac{\partial H}{\partial h},\label{pure_ad3}
\end{align} 
which have the natural boundary conditions
\begin{align}
\lambda_{\tilde{U}}(L)&=\frac{\partial \Phi}{\partial \tilde{U}}=0,\\
\lambda_p(L)&=\frac{\partial \Phi}{\partial p}=1.
\end{align}
There is no natural boundary condition for $\lambda_h$ since $h$ is already prescribed at $x=L$. Finally, as in Section \ref{small_shear}, the optimality condition is (\ref{ham_cases}). In order for there to be a singular arc, we must have $H_\alpha=0$ and $dH_\alpha/dx=0$ for a finite interval. Following Pitcher \cite{pitcher} and McDanell \& Powers \cite{mcdanell1971necessary}, in this interval the value of the control is given by $\alpha=\alpha^*$, where $\alpha^*$ is defined by
\beq
\frac{d^2H_\alpha}{dx^2}(\alpha^*)=0.\label{opt}
\eeq
Setting $H_\alpha=dH_\alpha/dx=0$, and using Eq. (\ref{opt}), we find that there will only be a singular arc when
\beq
\alpha^*= 
\frac{2 S \tilde{U}^2 (8 S \tilde{U}^2 - f)}{
3 f + f \tilde{U} + 8 S \tilde{U}^3} \quad \mathrm{for}\quad x \in[x_1,x_2],\label{alpha_sing}
\eeq
for some $x_2>x_1$. 
Thus, (\ref{ham_cases}) becomes
\beq
\alpha(x)=\begin{cases}
    \alpha_{max},  & \text{if } H_\alpha>0, \\
    \alpha^*,  & \text{if } H_\alpha=0, \\
   \alpha_{min},  & \text{if } H_\alpha<0, \\
\end{cases}\label{ham_cases_pure}
\eeq
and we can solve the coupled system (\ref{dyn1})-(\ref{ham_cases_pure}) numerically for the optimal control and corresponding solution. In certain cases, where the singular arc value $\alpha^*$ is constant, we find analytical solutions, which we discuss at the end of this section.

In Fig. \ref{results3}b, d, f we plot the solution to the optimal control problem found using the Hamiltonian approach over the solution found using the numerical optimisation routine outlined in Section \ref{numerical}. It is clear that both approaches have found the same solution, with the singular arc lying between $x_1/h_0=2.9$ and $x_2/h_0=32.9$. The singular arc represents the balance between mixing and widening effects in the diffuser. It is necessary to mix the non-uniform flow before widening it because widening tends to accentuate the non-uniform profile, producing a high-kinetic energy low-pressure outlet. Therefore, the control initially takes its minimum value $\alpha=0^\circ$. However, a straight section which is too long is detrimental to pressure recovery because of wall drag. Hence a widening section is required after a certain critical length. The optimum value of $\alpha$ in the widening section represents a balance between mixing and widening the flow. If $\alpha$ is too large, then the flow profile will become too non-uniform at the outlet. Conversely, if $\alpha$ is too shallow, wall drag losses are enhanced. The singular arc is interesting from both a mathematical point of view, but also from an engineering point of view. It clearly shows that diffuser designs for non-uniform inflow should take into account the nature of the non-uniform inflow profile. The optimum widening angle for manufacture is given by Eq. (\ref{alpha_sing}). Unfortunately, Eq. (\ref{alpha_sing}) is difficult to calculate in general ($\tilde{U}$ is a variable), but we find that for certain parameter values, it takes a simpler and more useful form.

\begin{figure}
\centering
\centering
\begin{overpic}[width=0.45\textwidth]{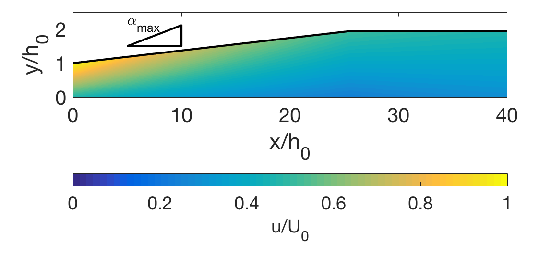}
\put(0,45){(a)}
\end{overpic}
\begin{overpic}[width=0.45\textwidth]{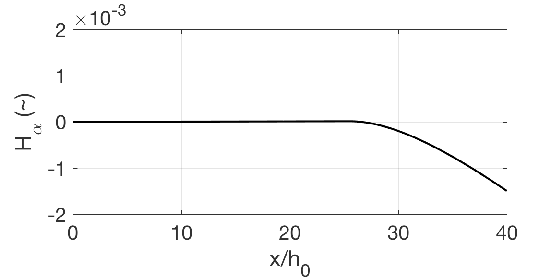}
\put(0,45){(b)}
\end{overpic}
\caption{Constant singular arc solution displaying velocity colour map and a plot of $H_\alpha=\partial H/\partial \alpha$, where the Hamiltonian $H$ is given by (\ref{pure_ham}), for parameter values $S=0.11$ and $f=0.01$. The inflow velocity ratio $U(0)=0.471$ is given by (\ref{U_eq}) and the singular arc value $\alpha^*=\tan 2.26^\circ $ is given by (\ref{alpha3}).\label{pure_sing_arc}}
\end{figure}

In general, the value of $\alpha$ during the singular arc is not constant, yet in certain cases, such as Fig. \ref{results2}b, it doesn't vary much over the singular arc interval. This raises the question of whether it is possible to find constant $\alpha^*$ solutions. Noticing how the only variable in Eq. (\ref{alpha_sing}) is $\tilde{U}$, we seek solutions with constant $\tilde{U}$. Furthermore, we restrict our attention to solutions which begin on the singular arc. From Eq. (\ref{dyn1}), we see that constant $\alpha^*$ solutions only exist if
\beq
\alpha^*=S\tilde{U}.\label{alpha2}
\eeq
Therefore, reconciling Eqs. (\ref{alpha_sing}) and (\ref{alpha2}), it can be shown that constant singular arc solutions exist for parameters which satisfy
\beq
S \tilde{U} (8 S \tilde{U}^3 - 3 f(1+ \tilde{U} ) )=0.\label{params}
\eeq
Excluding $\tilde{U}=0$, and assuming that $f$ and $S$ are fixed, we are left with solving Eq. (\ref{params}) for $\tilde{U}$. Substituting $U_2(0)$, $U_0$ and $h_0$ back into (\ref{params}), we rewrite the equation in terms of the inlet velocity ratio $U(0)=U_2(0)/U_0$, giving
\beq
4 S (U(0) -1 )^3 + 3 f (U(0)+ 1 )^2=0.\label{U_eq}
\eeq
Similarly, (\ref{alpha2}) becomes
\beq
\alpha^*=S\lb\frac{1-U(0)}{1+U(0)}\rb.\label{alpha3}
\eeq
{Considering that $f/S\approx0.1$ is small, we expand (\ref{U_eq}) and (\ref{alpha3}) about $f/S$ and ignore imaginary solutions, giving the approximate solution
\begin{align}
U(0)&=1 - 3^{1/3}\lb\frac{f}{S}\rb^{1/3}+ \frac{1}{3^{1/3}}\lb\frac{f}{S}\rb^{2/3}+\ldots,\\
\frac{\alpha^*}{S}& = \frac{3^{1/3}}{2} \lb\frac{f}{S}\rb^{1/3}+  \frac{1}{4 \cdot3^{1/3}}\lb\frac{f}{S}\rb^{2/3}+\ldots.
\end{align}}
We also need to ensure that both $H_\alpha=0$ and $dH_\alpha/dx=0$ for all values of $x$. This will enforce further constraints on the other parameters of the problem (e.g. $h_L/h_0$ and $L/h_0$), which we do not discuss here. As an example of such a constant singular arc solution, we choose parameter values $S=0.11$ and $f=0.01$, from which, solving Eqs. (\ref{U_eq}) and (\ref{alpha3}), we have an inflow velocity ratio $U(0)=0.471$ and a singular arc value of $\alpha^*=\tan 2.26^\circ$. We find that $H_\alpha=0$ and $dH_\alpha/dx=0$ along the singular arc if we choose the remaining parameter values $h_L/h_0=1.94$ and $L/h_0=40$. In Fig. \ref{pure_sing_arc} we display the velocity colour map for this solution, together with a plot of $H_\alpha$. Clearly we see that the solution starts on the singular arc until the expansion ratio $h_L/h_0=1.94$ is reached, at which point $H_\alpha$ becomes negative, such that the remaining length of the diffuser has angle $\alpha_{min}=0^\circ$.

\begin{figure}
\centering
\includegraphics[width=0.5\textwidth]{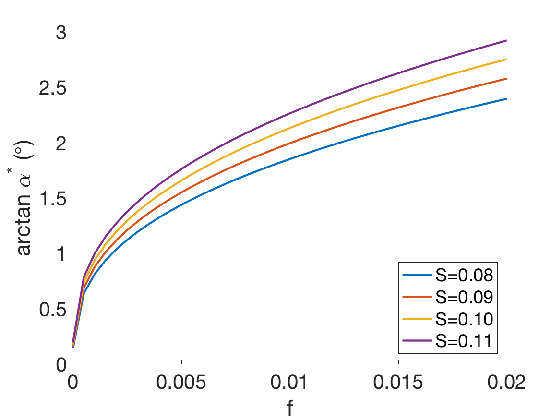}
\caption{Parameter analysis for the constant $\alpha^*$ singular arc solution,
given by (\ref{alpha3}), which depends on the parameters $S$ and $f$. \label{sing_arc}}
\end{figure}

It is of further interest to investigate how the singular arc depends on the model parameters $f$ and $S$. We plot the relationship between the constant singular arc value $\alpha^*$ (\ref{alpha3}) and these parameters in Fig. \ref{sing_arc}. It is clear that increasing the friction factor $f$ results in a higher $\alpha^*$. This is to be expected, since larger wall drag will penalise smaller angles more. Increasing the spreading parameter $S$ also increases $\alpha^*$. This is because higher spreading rates results in better mixing, and hence, wider angles are more affordable. 

\section{Comparison with results from a $k$-$\epsilon$ model and a $k$-$\omega$ SST model \label{cfd_comp_sec}}

In this section we discuss comparisons between the optimal shapes found using the simple model in the previous sections to calculations from CFD. Benham et al. \cite{benham2017turbulent} make comparisons between this model and a $k$-$\epsilon$ turbulence model \cite{launder1974numerical}, as well as experimental data generated with Particle Image Velocimetry (PIV). {Here we use both a $k$-$\epsilon$ and a $k$-$\omega$ Shear Stress Transport (SST) model \cite{menter1992improved} to compare with some of the simple model optimisation results. The $k$-$\epsilon$ model is one of the most popular computational turbulence models, whilst the $k$-$\omega$ SST model is particularly robust in situations with strong adverse pressure gradients \cite{menter1992improved} (though for the small diffuser angles we consider in this study the adverse pressure gradients are not severe).} Moreover, since a thorough comparison between the mathematical model and CFD has already been discussed in \cite{benham2017turbulent}, we do not perform comparisons for all of the optimisation results of Section \ref{numerical}. Instead we look at the geometry in Fig. \ref{results1}e as a single example.

Consider the example in Fig.  \ref{results1}e, as discussed in Section \ref{numerical}. In order to compare with the mathematical model we use precisely the same inlet velocity profile in the CFD. Inlet conditions for the turbulence variables $k$, $\epsilon$ and $\omega$ are given by the free-stream boundary conditions \cite{schlichting1960boundary} $k=I^2\times3/2\lb u^2+v^2\rb$, $\epsilon=0.09 k^{3/2}/\ell$ and $\omega=\sqrt{k}/\ell$, with turbulence intensity $I=10\%$ and mixing length $\ell=0.1 h_0$ ($10\%$ of the channel half-width). {In both the CFD models, no slip boundary conditions are applied to the channel walls. Furthermore, we use all the standard turbulence parameter values, which are given by Launder \& Spalding \cite{launder1974numerical} for the $k$-$\epsilon$ model and Menter \cite{menter1992improved} for the $k$-$\omega$ SST model.}

\begin{figure}
\centering
\vspace{0.5cm}
\centering
\begin{overpic}[width=0.45\textwidth]{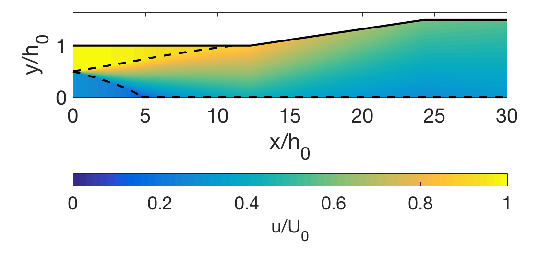}
\put(0,45){(a)}
\put(34,49){\bf Simple model}
\end{overpic}
\begin{overpic}[width=0.45\textwidth]{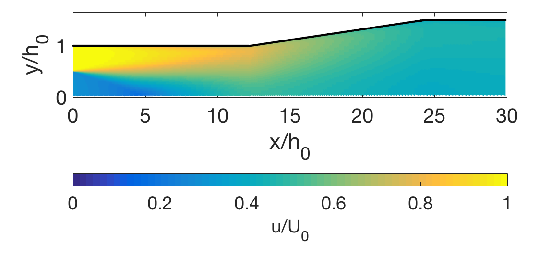}
\put(0,45){(b)}
\put(38,49){\bf $k$-$\epsilon$ model}
\end{overpic}\\
\begin{overpic}[width=0.45\textwidth]{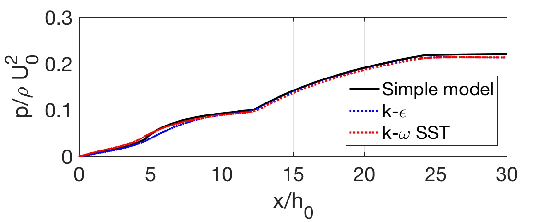}
\put(-5,38){(c)}
\end{overpic}
\begin{overpic}[width=0.45\textwidth]{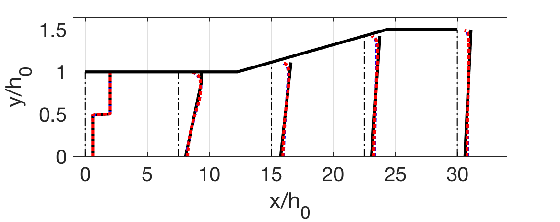}
\put(-5,38){(d)}
\end{overpic}
\caption{ 
Comparison between mathematical model and two computational turbulence models ($k$-$\epsilon$ and $k$-$\omega$ SST) for the optimal shape in Fig. \ref{results1}c. 
(a) Velocity colour map  calculated using the reduced model, with black dashed lines indicating the shear layer. (b) Corresponding velocity colour map calculated using the $k$-$\epsilon$ model.
(c) Pressure, averaged across the channel width. (d) Velocity profiles at evenly spaced locations in the channel.  \label{cfd_comp}}
\end{figure}

In Fig. \ref{cfd_comp}a, b we display colour plots of the time-averaged streamwise velocity $u$ generated with both the simple model and the $k$-$\epsilon$ model. {Figure \ref{cfd_comp}d also compares velocity profiles at evenly spaced locations in the channel, for the simple model and both CFD models.} There is good agreement between the models, with the simple model capturing the dominant features of the flow, such as maximum and minimum velocities, and the width of the shear layer. There is a slight discrepancy near the diffuser wall since our model does not resolve boundary layers, but instead parameterises their effect with a friction factor. However, we can see that our model accurately captures the effect of the boundary layers on the pressure by the close comparison between the models in Fig. \ref{cfd_comp}c.

In Section \ref{numerical}, we investigated reducing the dimension of the control $\alpha$ by splitting it into three piecewise constant sections divided by $x_1$ and $x_2$, which we treated as free parameters. 
Motivated by these piecewise linear shapes, here we make the same simplification, reducing the degrees of freedom of the control to $2$. 
We explore the parameter space generated by $x_1$ and $x_2$, using both CFD models to calculate pressure recovery. Each calculation made by the CFD models is much more computationally expensive than that of the simple model, but because of the low dimension of the degrees of freedom, we can still feasibly explore the different possible combinations of $x_1$ and $x_2$. This would not be tractable, however, if we were to use $100$ degrees of freedom, as we did with the simple model in Section \ref{numerical}. Hence, the numerical optimisation using the simple model is very useful for finding the general shape of the optimal channels, around which we can further search for optima using more realistic, yet more computationally intensive CFD models.

\begin{figure}
\centering
\centering
\begin{overpic}[width=0.45\textwidth]{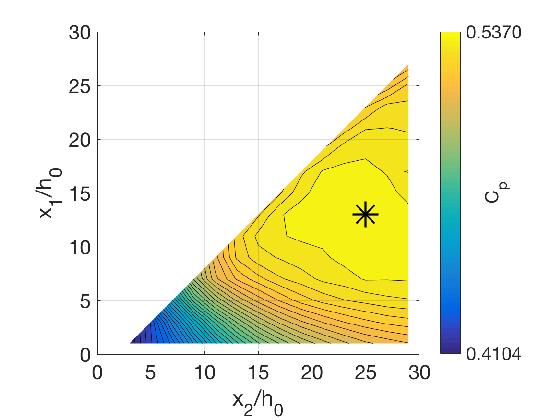}
\put(0,65){(a)}
\put(20,62){\bf Two-dimensional}
\end{overpic}
\centering
\begin{overpic}[width=0.45\textwidth]{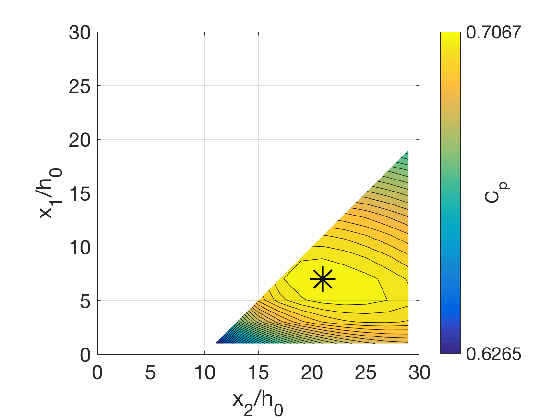}
\put(0,65){(b)}
\put(22,62){\bf Axisymmetric}
\end{overpic}
\caption{Contour plots of pressure recovery $C_p$, given by (\ref{C_p}), (\ref{C_p_axis}), over all permissible values of $x_1$ and $x_2$, calculated using the $k$-$\epsilon$ model. Direct comparison is made with Fig. \ref{results2}, where the same contour plots are calculated using the simple model. \label{contour_cfd}}
\end{figure}

In Fig. \ref{contour_cfd} we plot contours of pressure recovery $C_p$, given by Eq. (\ref{C_p}), as a function of the two parameters $x_1$ and $x_2$, where $C_p$ is calculated using the $k$-$\epsilon$ model instead of the simplified model, as in Fig. \ref{results2}. Similarly to the contour plots in Fig. \ref{results2}, we exclude values of $x_1$ and $x_2$ which result in a diffuser angle larger than $\tan 7^\circ$ for the two-dimensional case and $\tan 3.5^\circ$ for the axisymmetric case.

Comparing Figs. \ref{results2} and \ref{contour_cfd},  we see that the optimum diffuser shape using both the simple model and the $k$-$\epsilon$ model, is characterised by similar values of $x_1$ and $x_2$. In the two-dimensional case, according to the $k$-$\epsilon$ model, the optimum diffuser shape has $x_1/h_0=13$ and $x_2/h_0=25$, with a pressure recovery of $C_p=0.5370$. According to the simplified model, the optimum diffuser shape has $x_1/h_0=12.3$ and $x_2/h_0=24.3$, with a pressure recovery of $C_p=0.5205$, which is very close to that obtained with the $k$-$\epsilon$ model. Similarly, for the axisymmetric case, the $k$-$\epsilon$ model suggests an optimum diffuser shape with $x_1/h_0=7$ and $x_2/h_0=21$, giving a pressure recovery of $C_p=0.7067$, whereas the simplified model suggests $x_1/h_0=6.3$ and $x_2/h_0=19.8$, with a pressure recovery of $C_p=0.6886$.  {Considering that the diffuser angle is given by Eq. (\ref{wide_angle}), it is clear that if the optimum values of $x_1$ and $x_2$ are similar, according to the simple model and the CFD, then the optimum diffuser angle is also similar. In fact, we can compare the value of $C_p$ as a function of diffuser angle by looking at intersections of the contour plots (Figs. \ref{results2} and \ref{contour_cfd}) with the lines $x_1-x_2=\mathrm{const}$.}
{We have also generated these pressure recovery data using the $k-\omega$ SST model and we find the results very similar. The average discrepancy between the $C_p$ values calculated using the $k$-$\omega$ SST model and the $k$-$\epsilon$ model is $0.004$ for the two-dimensional case, and $0.003$ for the axisymmetric case.} 

These results indicate that the optimal shapes found using the numerical optimisation routine and the simplified model are similar to the optimal shapes that would be found if we were to use either of these computational turbulence models as a forward model. Hence, this gives us confidence that the optimal shapes generated using the simplified model are close to true optimal shapes in reality.

\section{Discussion and conclusion \label{discuss}}
\subsection{The effect of parameter values on the optimal shape}

Although we have investigated optimum diffuser shapes in a number of specific cases, we have not yet explored the various parameters of the model thoroughly, which are listed in Table \ref{table}. We now briefly discuss the effect that each of these parameters has on the optimal shapes. However, since there are many parameters, we do not provide plots for the analysis of every single parameter.

\begin{figure}
\centering
\vspace{0.5cm}
\begin{overpic}[width=0.45\textwidth]{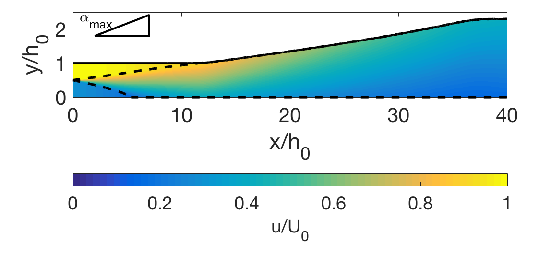}
\put(0,45){(a)}
\put(32,50){$\mathbf{U_2(0)/U_0=0.3}$}
\end{overpic}
\begin{overpic}[width=0.45\textwidth]{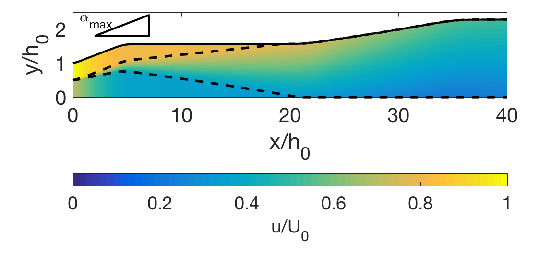}
\put(0,45){(b)}
\put(32,50){$\mathbf{U_2(0)/U_0=0.7}$}
\end{overpic}\\
\begin{overpic}[width=0.45\textwidth]{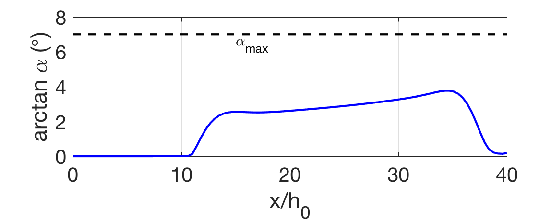}
\put(0,40){(c)}
\end{overpic}
\begin{overpic}[width=0.45\textwidth]{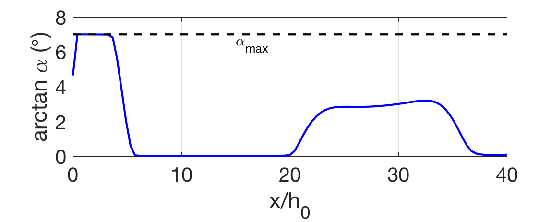}
\put(0,40){(d)}
\end{overpic}
\caption{Investigation of the dependence of optimal diffuser shapes on the inflow velocity ratio in the two-dimensional case, showing velocity colour maps (a, b) and plots of the control $\alpha$ (c, d). In both cases we choose an inflow with $h_2(0)/h_0= 0.5$, an expansion ratio $h_L/h_0 = 2.3$ and a length ratio $L/h_0 = 40$. The lower limit for the diffuser angle in both cases is $\alpha_{min}=0^\circ$. The upper limit is $\alpha_{max}= \tan 7^\circ$. \label{extraplts}}
\end{figure}

One of the most important parameters is the velocity ratio $U_2(0)/U_0$ of the inflow. To explore this parameter, we investigate optimal diffuser shapes for a fixed inflow with $h_2(0)/h_0 = 0.5$, an expansion ratio $h_L/h_0 = 2.3$ and a length ratio $L/h_0 = 40$, and we vary the velocity ratio. The results of the optimisation are displayed in Fig. \ref{extraplts}, for $U_2(0)/U_0=0.3$ and $U_2(0)/U_0=0.7$. We see that the effect of a velocity ratio which is closer to $1$ is that an initial widening section becomes favourable. This is because when the inflow is more uniform, wider angles penalise pressure recovery less. Therefore, the balance is tipped in favour of reducing wall drag by expanding the channel a little. In the extreme case where $U_2(0)/U_0$ becomes close to unity, we have seen in Section \ref{small_shear} that this initial widening section dominates throughout, such that the optimal control is purely bang-bang, with no singular arc. Notice how in the case of $U_2(0)/U_0=0.7$ the shape can no longer be approximated with the parameterisation of $x_1$ and $x_2$.

The effect of increasing or decreasing the ratio of the size of the plug regions from $h_2(0)/h_0=0.5$ is that the distance it takes for one of the plug regions to disappear becomes smaller. If the velocity ratio is small then, as described earlier, this distance is critical because it marks the point where the flow is sufficiently mixed that it is acceptable to expand thereafter at a wider angle. Hence, the effect of increasing or decreasing $h_2(0)/h_0$ is that this critical distance becomes smaller.

The effects of varying the diffuser expansion ratio $h_L/h_0$ and length ratio $L/h_0$ are more obvious and less interesting. Neither of them affect the optimum widening angle, but instead simply make the diffuser continue to expand wider and longer respectively. This is because they don't affect the crucial balance between wall drag and mixing effects. 

Varying the upper and lower bounds on the diffuser angle, $\alpha_{max}$ and $\alpha_{min}$, only affects the optimal solution if the diffuser angle touches the bounds over an interval. For example, in Fig. \ref{extraplts}a, c we see that $\alpha$ never touches $\alpha_{max}$. Therefore, in this case, raising $\alpha_{max}$ would have no effect on the solution. However, $\alpha$ clearly lies on the lower bound $\alpha_{min}$ over an interval at the beginning and near the end of the domain. Therefore, varying $\alpha_{min}$ here moves the optimal control along with it.


{Throughout this manuscript we have used a constant value of the spreading parameter $S=0.11$. In Section \ref{cfd_comp_sec} we showed that this parameter value is consistent with both a $k$-$\epsilon$ and a $k$-$\omega$ SST computational turbulence model using all standard turbulence parameter values \cite{launder1974numerical,menter1992improved}. In earlier work \cite{benham2017turbulent} we compared our simple model to PIV experiments in a three-dimensional geometry and found $S=0.18$. The equivalent spreading parameter for free shear layers \cite{pope2000turbulent} has been reported to take a range of values for different experiments and CFD calculations.}
Since the spreading parameter $S$ is associated with shear layer growth rate, for larger $S$, the shear layer will entrain the plug regions over a shorter distance. Similarly to varying $h_2(0)/h_0$, this decreases the critical distance after which expansion occurs. We have already discussed the effect of $S$ on the singular arc in Section \ref{pure_shear}.

For rougher channels with a larger friction factor $f$, thinner channels and smaller angles will be penalised more. In such cases, the optimum widening angle is larger. Furthermore, if $f$ is sufficiently large, it becomes more advantageous to have an initial widening section, similar to situations where the velocity ratio $U_2(0)/U_0$ is close to $1$. In the extreme case where wall drag dominates, the control becomes bang-bang because the penalty of worsening the non-uniform flow is eclipsed by the effect of wall drag. 
For very small wall drag, the optimal diffuser shape appears to prioritise mixing over widening the flow.
In these cases, thin channels and small diffuser angles are not penalised very much, such that the
critical distance after which expansion occurs 
is precisely the point where the shear layer has reached across the entire channel. At this point the flow is sufficiently mixed and can afford expansion.

\subsection{Conclusions}

We have developed a numerical optimisation routine to find the diffuser shape which maximises pressure recovery for given non-uniform inflow, in both two-dimensional and axisymmetric cases. The optimisation uses a simplified mathematical model for the development of turbulent shear layers in confining channels. We find that some of the optimal diffuser shapes are well approximated by shapes which are composed of two straight sections separated by a widening section with a constant widening angle. This is in contrast to diffuser design for uniform flow, where diffusers do not typically have an initial straight section. Furthermore, we show that the optimum widening angle is less than the angle at which boundary layer separation typically occurs, which is usually the diffuser angle chosen for uniform flow. Therefore, we have shown that the effects of non-uniform inflow are critical to diffuser performance, and should not be ignored when it comes to diffuser design.

In two limiting cases we use analytical techniques to interpret the optimal diffuser shapes found with the numerical optimisation. The first of these cases is the small shear limit, where the inflow is almost uniform, in which case the optimal control is bang-bang, such that the diffuser widens at the maximum possible angle until it reaches the desired cross-sectional area, and then remains at that area. The second case is the pure shear limit, where the inflow is a purely sheared flow with no plug regions. In this case, the optimal control may have a singular arc where, on an interval, the diffuser angle takes values between its upper and lower bounds. We show that in certain cases the singular arc corresponds to a constant angle and this angle depends on the friction factor $f$ and the spreading parameter $S$. We compare some of the numerical optimisation results with CFD simulations using both a $k$-$\epsilon$ and a $k$-$\omega$ SST turbulence model {(using all standard turbulence parameter values)}, finding good agreement. In the case where we approximate the diffuser shape with piecewise linear sections, we show that both the simplified model and the CFD  share almost identical optimal shapes. This suggests that the optimal shapes found using the numerical optimisation and the simplified model are indeed close to the true optimal shapes in reality.

\appendix
\section{Diffuser stagnation}\label{appA}

We have discussed how diffusers have the tendency to accentuate non-uniform flow. In extreme cases, where the diffuser angle is too large, it is possible for regions of the flow to slow to zero velocity and recirculate. Let us now consider diffusers which have a non-uniform inflow velocity given by (\ref{piecewise}). In our mathematical model, outlined in Section \ref{model_model}, we account for the possibility of a stagnated region by introducing the velocity of the slower plug region $U_2$ into Bernoulli's equation (\ref{bernoulli_stag2}) in the form of a complementarity condition. In this way, if the slower central plug region reaches zero velocity, we model this as a region of dead water and maintain it at zero velocity. 
In reality these regions have relatively slow recirculation. We do not resolve the recirculation, but instead we resolve the size of the regions and treat them as having average streamwise velocity $u=0$. In order to justify this model assumption we compare it with CFD calculations for a diffuser with a stagnated region.

\begin{figure}
\centering
\begin{overpic}[width=0.45\textwidth]{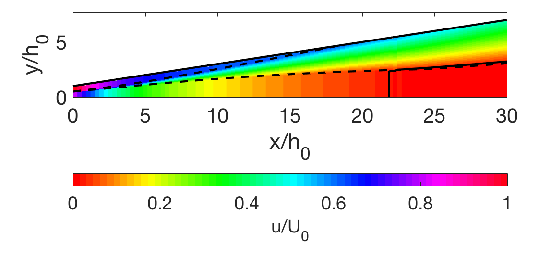}
\put(0,45){(a)}
\end{overpic}
\begin{overpic}[width=0.45\textwidth]{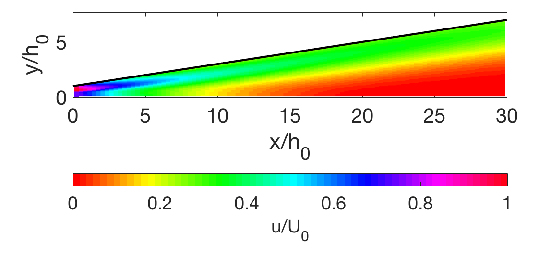}
\put(0,45){(b)}
\end{overpic}\\
\begin{overpic}[width=0.45\textwidth]{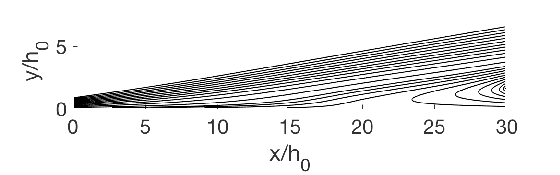}
\put(0,40){(c)}
\end{overpic}
\begin{overpic}[width=0.45\textwidth]{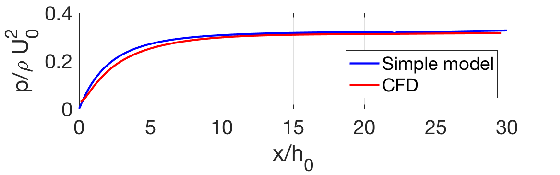}
\put(0,40){(d)}
\end{overpic}
\caption{Comparison between our simple model and a $k$-$\epsilon$ CFD model for a two-dimensional diffuser with a stagnation region in the centre. (a) Velocity colour map  calculated using the reduced model, with black dashed lines indicating the shear layer, and a solid black contour indicating the stagnated zone. (b) Corresponding velocity colour map calculated using  the $k$-$\epsilon$ model. (c) Streamlines calculated from the $k$-$\epsilon$ model. (d) Pressure profile averaged across the width of the channel. An alternative colour scheme is used in the colour maps for the purposes of illustrating regions of zero velocity clearly. \label{stag}}
\end{figure}

To make the comparison, we choose an inflow velocity with $U_2(0)/U_0=0.75$ and $h_2(0)/h_0=0.5$. We choose a value of $U_2(0)/U_0$ fairly close to $1$ to show that stagnation can occur for even moderately non-uniform inflow conditions. The diffuser that we select has constant widening angle $\alpha=\tan 11^\circ$ and has non-dimensional length $L/h_0=30$. For the CFD, we use the same $k$-$\epsilon$ turbulence model as in Section \ref{cfd_comp_sec}. In Fig. \ref{stag} we compare the results of the model and the CFD. Time-averaged velocity colour maps are compared in Fig. \ref{stag}a, b, where we indicate the stagnated region in our model (a) with a black contour. The comparison is good, with the model capturing the dominant flow features, such as the width of the shear layer. In Fig. \ref{stag}c we display streamlines calculated using CFD. These indicate that there is indeed a stagnated region with recirculation in the centre of the diffuser. This region is located in approximately the same position, and has approximately the same size as the prediction from the mathematical model. The time-averaged pressure profile, averaged across the channel width is plotted in Fig. \ref{stag}d and, again, it shows good agreement between the CFD and our model, suggesting that our model accurately captures the general behaviour of the diffuser when it has a stagnated region.

\begin{acknowledgements}
This publication is based on work supported by the EPSRC Centre for Doctoral Training in Industrially Focused Mathematical Modelling (EP/L015803/1) in collaboration with VerdErg Renewable Energy Limited and inspired by their novel Venturi-Enhanced Turbine Technology for low-head hydropower.
\end{acknowledgements}

\bibliographystyle{/Users/Graham/Documents/Maths/verderg/papers/Diffuser_shape_optimal_control/JMathEng/Diffuser_control_v3/cj}
\bibliography{testbib.bib}


\end{document}